\documentclass[final,onefignum,onetabnum]{siamltex1213}
\usepackage{amsmath}
\usepackage{amsfonts}
\usepackage{newfloat}

\newcommand{\algcom}[1]{\hfill\(\triangleright\) {\footnotesize #1}}
\DeclareMathOperator*{\argmin}{arg\,min}
\DeclareMathOperator{\tree}{tree}
\DeclareMathOperator{\children}{children}
\DeclareMathOperator{\Var}{Var}
\newcommand{\app}[1]{{#1}^\star}
\newcommand{\tapp}[1]{\hat{#1}^\star}
\newcommand{\prim}[1]{\widebar{#1}}
\newcommand{\seco}[1]{\tilde{#1}}
\newlength\figwidtht
\newlength\figwidthe
\newlength\figwidthes
\DeclareFloatingEnvironment[fileext=loz,listname={List of Algorithms},name=Algorithm,placement=bpt,within=none]{zalg}
\makeatletter
\newcommand*\rel@kern[1]{\kern#1\dimexpr\macc@kerna}
\newcommand*\widebar[1]{%
  \begingroup
  \def\mathaccent##1##2{%
    \rel@kern{0.8}%
    \overline{\rel@kern{-0.8}\macc@nucleus\rel@kern{0.2}}%
    \rel@kern{-0.2}%
  }%
  \macc@depth\@ne
  \let\math@bgroup\@empty \let\math@egroup\macc@set@skewchar
  \mathsurround\z@ \frozen@everymath{\mathgroup\macc@group\relax}%
  \macc@set@skewchar\relax
  \let\mathaccentV\macc@nested@a
  \macc@nested@a\relax111{#1}%
  \endgroup
}
\makeatother

\begin{document}

\renewcommand{\thefootnote}{\fnsymbol{footnote}}

\title{Compressive sampling for \\ energy spectrum estimation of turbulent flows\footnotemark[1]} 

\author{Gudmundur F. Adalsteinsson\footnotemark[2] \and Nicholas K.-R. Kevlahan\footnotemark[3]}

\footnotetext[2]{School of Computational Science and Engineering, McMaster University, Hamilton, ON L8S 4K1, Canada}
\footnotetext[3]{Department of Mathematics and Statistics, McMaster University, Hamilton, ON L8S 4K1, Canada}

\footnotetext[1]{Submitted to the SIAM Journal on Scientific Computing on April 22, 2014.} 

\renewcommand{\thefootnote}{\arabic{footnote}}

\slugger{sisc}{2014}{xx}{x}{x--x}
\maketitle

\pagestyle{myheadings}
\thispagestyle{plain}
\markboth{G. F. ADALSTEINSSON AND N. K.-R. KEVLAHAN}{COMPRESSIVE SAMPLING FOR SPECTRUM ESTIMATION} 

\begin{abstract}
Recent results from compressive sampling (CS) have demonstrated that accurate reconstruction of sparse signals often requires far fewer samples than suggested by the classical Nyquist--Shannon sampling theorem. Typically, signal reconstruction errors are measured in the $\ell^2$ norm and the signal is assumed to be sparse, compressible or having a prior distribution. Our spectrum estimation by sparse optimization (SpESO) method uses prior information about isotropic homogeneous turbulent flows with power law energy spectra and applies the methods of CS to 1-D and 2-D turbulence signals to estimate their energy spectra with small logarithmic errors. SpESO is distinct from existing energy spectrum estimation methods which are based on sparse support of the signal in Fourier space.   SpESO approximates energy spectra with an order of magnitude fewer samples than needed with Shannon sampling. Our results demonstrate that SpESO performs much better than lumped orthogonal matching pursuit (LOMP), and as well or better than wavelet-based best $M$-term or $M/2$-term methods, even though these methods require complete sampling of the signal before compression.  
\end{abstract}


\section{Introduction\label{sec:intro}} 
Sampling and storage of signals becomes challenging for high wavenumber or high dimensional signals if the Nyquist--Shannon sampling theorem is followed strictly. The theory of compressive sampling (CS) provides a rigorous framework to accurately reconstruct a signal from a few non-adaptive (random) projections, provided it is sufficiently sparse or compressible in some basis~\cite{robunpr,donoho2006cs,candes2006quantru}. Since statistically homogeneous turbulent signals are not known for their high compressibility, the use of CS for turbulence is on the edge of applicability.  In addition, turbulence researchers are often more interested in reconstructing Fourier energy spectra from spatial measurements and spectrum estimation is not a well-developed area of CS.

Consider the discrete signal $u\in \mathbb R^{n^d}$ of length $N=n^d$ in $d$ dimensions. The traditional fixed-rate sampling, hereafter referred to {\em Shannon sampling\/}, of $u$ is inefficient if the coefficients $\hat u$ of $u$ in an orthogonal basis are sufficiently compressible. Shannon sampling is especially wasteful if we are  interested only in a particular low dimensional property of the signal, such as the one-dimensional energy spectrum of a two- or three-dimensional data set.  

This paper focuses on the reconstruction of energy spectra of homogeneous iso\-tro\-pic turbulent flows from a minimal number of samples. A turbulent flow is characterized by a non-dimensional number, the Reynolds number Re, which is the ratio of inertial terms to viscous terms in the Navier--Stokes equations governing the flow.  Flows become turbulent when Re exceeds a certain threshold (typically $\sim\! 10^3$) and industrial and natural turbulent flows have very large Reynolds numbers ($\sim\! 10^5$--$10^{12}$). The minimum length scale of a turbulent flow, the Kolmogorov scale $\eta$, decreases with increasing Reynolds number $\text{Re}$  like $\eta\propto\text{Re}^{-3/4}$~\cite{frisch1999turbulence}, and the number of spatial samples required by the sampling theorem in $d$ dimensions is $N\propto\eta^{-d}$.  Therefore, the total number of samples needed to characterize a turbulent flow increases very quickly with Reynolds number: like $\text{Re}^{9/4}$ in three dimensions and $\text{Re}^{3/2}$ in two dimensions. Thus, straightforward application of Shannon sampling  requires huge amounts of regularly sampled data ($\sim\! 10^{11}$--$10^{27}$) to estimate the complete one-dimensional energy spectrum of a three-dimensional turbulent flow.  

However, because the range in wavenumber space of the one-dimensional energy spectrum of $u$ is proportional to $\eta^{-1} \propto \text{Re}^{3/4}$, there is definitely room for improved sampling strategies. Even for one-dimensional signals, such as hot-wire measurements, it should be possible to accurately characterize the energy spectrum using fewer samples than required for the usual Shannon sampling. 

In order to accurately estimate the one-dimensional energy spectra of signals with a very large and continuous range of active length scales, we propose a new method that uses {\em a priori\/} information about the signal, such as the structure and scaling of wavelet coefficients, isotropy, and power law behaviour of  the energy spectrum. We show that our method is able to approximate energy spectra with an order of magnitude fewer samples than needed with Shannon sampling.

We introduce notation and give a brief introduction to CS  in section \ref{sec:cs} before we define our problem and introduce two measurement matrix types used in our experiments.  In section \ref{sec:est} we introduce the relevant wavelet transforms and their application to turbulence, and finally present our SpESO algorithm for estimating energy spectra. Section~\ref{sec:num} verifies the method by applying it to a set of representative test cases: 1-D hot-wire turbulence data, 1-D synthetic power-law data,  2-D numerical simulation turbulence data and 2-D synthetic power-law signals.

In related work, variants of CS have been developed to estimate spectra and other properties of signals, but in different contexts which do not apply in our case.  In \cite{davenport2010signalpr} linear functions of signals were estimated by fast operators. Energy spectra, however, are nonlinear functions of signals. Sparse and locally supported 2-D spectra were estimated in \cite{venkataramani1998blind}, but turbulence is not sparse in Fourier space. Similarly, \cite{leus2011power,leus0011coset} put some sparsity constraints on their power spectrum estimation.  Bands of power spectra are estimated on a linear scale from non-uniform samples in \cite{lexa2011spectral}. In \cite{almeida2012spectra} the 2-D spectrum itself is sampled and approximated to reduce computational time in spectroscopy.  General nonlinear optimization problems for CS are considered in \cite{blumensath2012nonlinear}. However, the iterative algorithm proposed is impractical  in our case as it requires expensive high dimensional gradients to be computed at each iteration.

\section{Compressive sampling for large signals\label{sec:cs}} 
In this paper we assume that the turbulent flow is provided as a single component of a turbulent velocity vector field as a discrete sequence $u\in\mathbb R^N$. Mathematically, of course, the flow is more accurately described as velocity (or vorticity) vector field of velocity defined on a three-dimensional spatial domain. However, assuming the flow is band-limited in wavenumber, the Nyquist--Shannon sampling theorem allows us to represent it as sequence of discrete values. The measurement matrices discussed later are discrete approximations of linear operators in continuous space.  We represent two-dimensional signals of dimension $n\times n$ as vectors of length $N=n^2$.

We first decompose $u$ as a linear combination of vectors in a basis $\Phi\in\mathbb R^{N\times N}$ ,
\begin{equation}
	u=\Phi\hat u=\sum_{i} \hat u_i\phi_i,
\end{equation}
where $\hat u$ are the expansion coefficients and $\phi_i$ are the basis vectors. A signal $u$ is said to be $B$\emph{-sparse} in basis $\Phi$ if $|\supp(\hat u)|=B<N$, where $|\cdot |$ denotes cardinality and $\supp(x)=\{ i : x_i\neq0\} $ is the support. 

The signal $u$ is called \emph{compressible\/} in the basis $\Phi$ if it has ordered coefficients $|\hat u|_{(1)}\geq \dots \geq |\hat u|_{(N)}$ that satisfy the inequality $|\hat u|_{(n)} \leq C n^{-s}$ for $s>0$ and a constant $C$~\cite{candes2005practical}. The \emph{best $B$-term approximation} in an orthonormal basis, $f_B$, is an approximation with all but the $B$ largest terms of $\hat u$ zero. Many signals are highly compressible in a wavelet basis \cite{vetterli1998datacompression} since wavelet basis functions are self-similar and are localized in both position and scale. If the signal is compressible then the error in the best $B$-term approximation is $\| f_B -f\|=\mathcal O(B^{-s+1/2})$.

The central idea of CS, see e.g.\ \cite{robunpr,donoho2006cs,candes2006quantru,candes2006stable}, is that a few linear non-adaptive (e.g. random) measurements of a signal are sufficient to accurately reconstruct a signal if that signal is compressible in some basis. Note that the measurement scheme (e.g. random samples) and the sparsity system (e.g. a wavelet basis) must be mutually incoherent in the sense of having a sufficiently small maximum inner product between the basis vectors of the measurement scheme and the sparsity system.

Let $A\in\mathbb R^{M\times N}$ be a \emph{measurement matrix\/}, let $g\in \mathbb R^M$ be the \emph{compressed samples\/}, and assume $M<N$. The measurement scheme is defined by the under-determined system
\begin{equation}\label{eq:cssys}
	g=Au.
\end{equation}
In a slightly different form, with $\Psi=A\Phi$ which we call the \emph{CS-matrix}, we have
\begin{equation} \label{eq:cssyssp}
	g=\Psi \hat u,
\end{equation}
where $\hat u$ is assumed to be $B$-sparse in the basis $\Phi$. Under this framework, the minimization problem \cite{candes2006stable} 
\begin{equation} \label{eq:bp}
	\tapp u = \argmin_{\substack{\hat h\in\mathbb R^N }} \| \hat h\|_{\ell^1}
	\quad\text{s.t.}\;\;  \Psi\hat h=g,
\end{equation}
is proved to accurately approximate, or exactly reconstruct, the original signal, provided some basic conditions on the structure of $\Psi$ and the compressibility of the signal are satisfied.  (A star superscript, $\app u$, denotes approximation.) This method is called {\em basis pursuit\/} and can be solved via convex optimization. Unfortunately, turbulent signals are not compressible enough in wavelet bases for basis pursuit to give meaningful results, especially in the high wavenumber range of the spectrum.

Reconstruction methods which are significantly faster than the basis pursuit method for \eqref{eq:bp} include so-called greedy methods. A popular greedy method is iterative \emph{orthogonal matching pursuit} (OMP) \cite{tropp2007signalrecovery}. Our estimation algorithm relies heavily on a multi-level modification of OMP called QOMOMP, see section \ref{sec:est}. OMP can be  generalized easily to estimate more than one coefficient of the signal at a time \cite{wang2012gomp}. The experiments in section \ref{sec:num} use Lumped OMP (LOMP) as a comparison to our SpESO method, where the sparsity $B_0$ is fixed and $L_0$ coefficients are estimated in each iteration, requiring a total of $B_0/L_0$ iterations.

The initial CS literature was largely concerned with full random measurement matrices $A$, which require $\mathcal O(NM)$ operations to apply to a vector.  Many CS decoding methods require frequent application of $A$ and its transpose. For very large signals the matrix--vector multiplications are very memory and CPU intensive \cite{candes2006sparsityand}, so a full random matrix is not practical.  In our method we consider two matrices with fast matrix-free transforms requiring at most $ \mathcal O(N \log N)$ operations and $\mathcal O(N)$ memory to apply.

The first matrix is intended for measurements of 1-D time-dependent signals---such as hot-wire measurements---without requiring the whole signal for every compressed sample: a random finite impulse response (FIR) \emph{filter} \cite{tropp2006randfilt}.  Let the filter coefficients $h$ be compactly supported with support size $K$. We can then write
\begin{equation}
 A=R_{\Gamma} F^{*}\Sigma F,
\end{equation}
where $\Sigma=\diag(F h)$ is a diagonal matrix where the diagonal elements are the Fourier transform of $h$, and $F$ is the Fourier transform matrix. Here $R_{\Gamma}$ restricts the result to an evenly distributed set $\Gamma$ of length $M$. This definition of $A$ assumes periodicity, but our implementation zero pads the signal before the convolution to account for non-periodic boundary conditions. For a downsampling fraction $\delta_0$ and with $1/\delta_0\in \mathbb N$, the number samples is 
\begin{equation} \label{eq:filterm}
M=\lceil (N+K-3)\delta_0\rceil,
\end{equation}
and the complexity is $\mathcal O(KM)$.

A \emph{random convolution} and sub-sampling is a universal sampling strategy \cite{romberg2009randconv}. Consider now a full vector $h$ and a diagonal matrix $\Sigma=\diag(h)$ which randomizes the phase, i.e.\ $h_k=e^{i \theta_k}$, where $\theta_k$ are i.i.d.\ uniformly distributed on $(0,2 \pi)$ such that $F^{*}\Sigma\in\mathbb R^N$. We can again write
\begin{equation}
 A=R_{\Gamma} F^{*}\Sigma F,
\end{equation}
where $R$ restricts the result to a random set $\Gamma$. The complexity of this approach is $\mathcal O(N\log N)$. Note that the random convolution matrix $A$ has the property that its right pseudo-inverse is the transpose, $A^T(A A^T)^{-1}= A^T$ (i.e. $A A^T = I$ or $A$ is right-orthogonal). We use this matrix (or measurement scheme) for analyzing 2-D data.

\section{Energy spectrum estimation of turbulence data}
\subsection{Problem formulation}
Our problem is challenging because we seek to estimate the energy spectrum $E(k)$ from measurements of $u$, rather than estimating $u$ directly.  This problem is challenging because the quantity to be estimated, $E(k)$, is a nonlinear function of the quantity that is sampled, $u$. In addition, $u$ is not sparse in Fourier space. Dropping the constant normalization factor, let us define $E(k)$ as
\begin{align} 
E(k) = \sum_{k\leq |\textbf{k}'| <k+1} \big|\hat u[\textbf{k}'] \big|^2,
\end{align} 
where $\hat u$ is the Fourier transform of $u$, and we use the convention that $E_f$ is the spectrum of signal $f$. Our problem can be stated in general terms as
\begin{equation} \label{eq:problem}
\min_{\app u\in \mathcal U} \| \log( E_{ u}) - \log( E_{ \app u}) \|_{w},
\end{equation}
where the solution has properties specified by $\mathcal U\subseteq \mathbb R^N$, and the $\ell^2$ norm has weight $w$. Obviously, the term $\log( E_{ u})$ is an unknown in \eqref{eq:problem}.  Since we have the samples $g=Au$ we can project the signals and recast the problem as
\begin{equation} \label{eq:ataproblem}
\min_{\app u\in \mathcal U} \| \log( E_{ A^T Au}) - \log( E_{ A^T A\app u}) \|_{w},
\end{equation}
which can be solved using available data. This is specific case of the general nonlinear minimization $\min_{x\in \mathcal U} f(x)$. There exists a gradient method for this problem with an iterative solver \cite{blumensath2012nonlinear}
\begin{equation}
x^{k+1} = \mathcal P_{\mathcal U} \left( x^k - \mu \nabla f(x^k)\right).
\end{equation}
However, this method has the drawback that the computation of the gradient of $f$ is very expensive. In section~\ref{sec:est} we introduce a more efficient method based on OMP to solve this key problem in energy spectrum estimation.

\subsection{Estimation algorithm\label{sec:est}} 
We now introduce our sparsity system,  the orthogonal discrete wavelet transform (DWT) \cite{daubechies1988ortho,mallat1989multi}.  We choose the DWT because many signals are compressible in a wavelet basis, and the properties of the wavelet transform of turbulence signals are well-known \cite{Schneider/Vasilyev:2010,farge1996wave}. The following wavelet analysis assumes a signal of size $2^J$, an integer power of two, with $0\leq j < J$. A full wavelet decomposition of a signal on $J$ scales is
\begin{equation}
u =  \hat s_0^0 \phi_0^0+\sum_{j=0}^{J-1} \sum_{i=0}^{2^{j}-1} \hat d_i^j \psi_i^j,
\end{equation}
where $\phi_i^j$ and $\psi_i^j$ are respectively the scaling and the wavelet functions and $\hat s_i^j$ and $\hat d_i^j$ are the expansion coefficients. The level is $j$, the scale is $2^{-j}$, and $i$ is the translation. For simplicity we assume a full transform with a single (coarse) scaling coefficient $\hat s_0^0$. Note that the basis of scaling functions $\{\phi^j\}$ span the approximation subspace $V^j$, while the wavelet basis spans $\{\psi^j\}$ spans the subspace $W^j$ which is the orthogonal complement of $V^j$ in $V^{j+1}$, i.e. $V^{j+1} = V^j \oplus W^j$. Thus, a wavelet coefficient $\hat d_i^j$ measurements the how big the signal variation is at a position $i$ and scale $j$.  

The DWT has a fast transform for discrete signals, with complexity $\mathcal O(N)$. The decay rate of the wavelet coefficients is determined by the local regularity of the signal \cite{mallat2009wavelet}, and this decay rate can be used to estimate the strength of any \mbox{(quasi-)}singularities in signal.  The coefficients $\hat s_i^j$ and $\hat d_i^j$ are stored in $\hat u$ in the standard manner. 

In 2-D a separable multi-resolution analysis (MRA) includes three components of scale variation \cite{mallat2009wavelet}, decomposing a $2^J\times 2^J$ signal similarly into
\begin{equation} \label{eq:wavelet2-D}
u =  \hat s_{0,0} \phi_{0,0} + \sum_{k=1}^{3}\sum_{j=0}^{J-1} \sum_{i_1,i_2=0}^{2^{j}-1} \hat d_{j,i}^k \psi_{j,i}^k 
\end{equation}
with $i=(i_1,i_2)$ and $k=1,2,3$ includes contributions from wavelets measuring variation in the horizontal, vertical and diagonal directions.

\begin{zalg} 
\caption{Quasi-Oracle Multilevel Orthogonal Matching Pursuit (QOMOMP). Approximates a sparse solution to $g=\Psi \hat u$, where $g\in \mathbb R^M$ and $\hat u\in \mathbb R^N$. It assumes $\hat u$ is in a wavelet basis and approximates all coefficients at levels $j<J_0$ and $L_j$ coefficients at level $j\geq J_0$. \label{alg:qomomp}}\small\normalfont
\vspace{-1mm} 
\hrule
\vspace{1mm} 
\hspace*{5mm} $\Omega \gets \cup_{j<J_0} \Gamma_j$  	\algcom{initial coefficient index set by oracle, $\Gamma_j$ is the index set for level $j$} \\
\hspace*{5mm} $\tapp u \gets 0$ 			\algcom{the decoded signal initial guess} \\
\hspace*{5mm} $\tapp u_{\Omega} \gets \argmin_{x} \|\Psi_{\Omega} x-g\|^2 $  	\algcom{least squares, $\Psi_{\Omega}$ are columns of $\Psi$} \\
\hspace*{5mm} \textbf{for} $j = J_0$ to $J-1$ \textbf{do}	\algcom{for each level $j\geq J_0$} \\
\hspace*{10mm} $r\gets g-\Psi \tapp u$		\algcom{update residual} \\
\hspace*{10mm} $a \gets \Psi^T r$			\algcom{project residual}\\
\hspace*{10mm} $i \gets \supp \big( |\tree(a)|_{L_j}^{>} \big)$	\algcom{the largest $L_j$ coefficients of $\tree(a)$ in $\Gamma_j$, see Algorithm \ref{alg:tree}} \\
\hspace*{10mm} $\Omega \gets \Omega\cup i$	\algcom{update current index set} \\
\hspace*{10mm} $x_0 \gets \tapp u_{\Omega}$	\algcom{initial guess} \\
\hspace*{10mm} $\tapp u_{\Omega} \gets \argmin_{x_0} \|\Psi_{\Omega} x_0- g\|^2$					\algcom{least squares} \\
\hspace*{5mm} \textbf{end for}					
\vspace{1mm} 
\hrule
\end{zalg}
% $a_{\Gamma_j^c} \gets 0$	

Turbulent flows have been analyzed and computed adaptively using wavelet methods for almost two decades \cite{farge1996wave}. Although homogeneous turbulence is not highly compressible,  wavelet coefficients approximate local structures much better than Fourier modes due to the intermittent multi-scale structure of turbulence. This multi-scale structure is characterized by a continuous range of active length scales which grows like $\text{Re}^{3/4}$ and has a power-law energy spectrum like $E(k)\propto k^{-5/3}$ in three dimensions.

The multi-scale structure of turbulence and the DWT leads us to propose a multi-level version of OMP that uses our knowledge about the multi-scale turbulent flows to predict the typical space and scale structure of the wavelet coefficients.   For example,  {\em a priori\/} we know that wavelet coefficients are relatively large above a certain scale and, on average, the magnitude of wavelet coefficients decreases monotonically with decreasing scale. We call this method \emph{quasi-oracle multilevel orthogonal matching pursuit} (QOMOMP), see Algorithm~\ref{alg:qomomp}.  QOMOMP will be used to efficiently solve the minimization problem (\ref{eq:bp}), which is the key computational step of our energy spectrum estimation method.

QOMOMP estimates all coefficients at levels less than a pre-defined coarsest level $j<J_0$. The ``initial coefficient index set by oracle'' defined by $J_0$ is chosen such that almost all wavelet coefficients up to level $J_0$ are large, approximately large enough to be included in the best $M/2$-term approximation. At each finer scale $j\geq J_0$ a pre-defined number of coefficients, $L_j$, is estimated. We will see later that the choice of the sequence $L=\{L_j\}$ is a key factor determining the performance of the method.

\begin{zalg} 
\caption{Description of the function $\tree(a)$. Returns adjusted elements of $a=\Psi^T r$ at level $j$ to enforce tree-like structure of the estimated coefficients $\tapp u_{\Omega}$. The function depends on the parameter $\beta\geq 1$ and a threshold defined by $\Lambda: \mathbb R^{|\omega|}\to\mathbb R_{+}$. \label{alg:tree}}\small\normalfont
\vspace{-1mm} 
\hrule
\vspace{1mm} 
\hspace*{5mm} $\omega \gets \Omega\cap\Gamma_{j-1}$  	\algcom{$\Omega$ is current coefficient index set, $\Gamma_{j-1}$ is index set for level $j-1$} \\
\hspace*{5mm} $\omega^\star \gets \{ i\in\omega : |\tapp u_i| > \Lambda (\tapp u_{\omega}) \}$ 	\algcom{locate large coefficients in $\Gamma_{j-1}$} \\
\hspace*{5mm} $\Omega^\star \gets \children(\omega^\star)$  \algcom{corresponding coefficients in $\Gamma_{j}$} \\
\hspace*{5mm} $a_{\Gamma_j^c} \gets 0$	\algcom{coefficients outside $\Gamma_j$ will not be selected} \\
\hspace*{5mm} $a_{\Omega^\star} \gets \beta a_{\Omega^\star}$	\algcom{adjust elements of $a$ with a large parent coefficient (in $\tapp u$)}
\vspace{1mm} 
\hrule
\end{zalg}

Discrete wavelet coefficients have a tree-like structure, where (in 1-D) the two child coefficients at a fine scale $j$ are more likely to be large if their parent coefficient at the coarse scale $j-1$ is large. To enforce this tree-like structure of the non-zero wavelet coefficients $\tapp u$ we apply the function $\tree(a)$, see Algorithm~\ref{alg:tree}, to modify the raw wavelet coefficients of the residual in the QOMOMP Algorithm~\ref{alg:qomomp}. This is similar to the method used in~\cite{la2005tree}, but enforces the tree structure less strictly. 

The tree algorithm~\ref{alg:tree} works as follows. Let $\Gamma_j$ be the index set for level $j$ and $\Omega$ be the current support of wavelet coefficients $\tapp u$ at iteration $j$ in QOMOMP.  The index set $\omega^\star$ identifies those coefficients at the coarse level $j-1$ above a threshold defined by $\Lambda$. Then, $\Omega^\star = \children(\omega^\star)$ are the child coefficients at level $j$ of the significant parent coefficients $\omega^*$ at level $j-1$. Finally, the tree function scales the residuals $a$ in $\Omega^\star$ by a constant, $a_{\Omega^\star} \gets \beta a_{\Omega^\star}$.  If $\beta>1$ this makes the residuals corresponding to children at scale $j$ of significant wavelet coefficients at scale $j-1$ more likely to be selected as the $L_j$ largest coefficients.  If $\beta=1$ $\tree(a)$ does nothing, while in the limit $\beta\rightarrow\infty$ it exactly enforces a tree structure.

Isotropy of the signal is not of concern in 1-D. In 2-D, however, the diagonal wavelet coefficients, denoted by $k=3$ in \eqref{eq:wavelet2-D}, of a best $B$-term approximation of an isotropic signal become a smaller proportion of the total for a particular level as the scale decreases. To account for this we let the operator $|\cdot |_{L_j}^{>}$ in QOMOMP in 2-D choose the coefficients such that the diagonal ones are a ratio $q_j$ of the total for level $j$.
\begin{figure}[tbp]
% cpu_scaling_plotload1.m
\begin{center}\footnotesize
\includegraphics[width=\figwidtht]{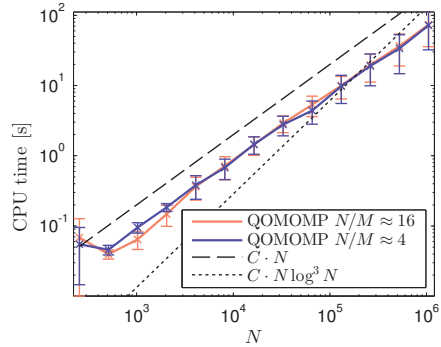}
\caption{Computational cost of QOMOMP (measured by CPU time) versus signal length $N$, showing mean curves and standard deviation bars of 16 random simulations. The number of samples is a fixed ratio of $N$, either $N/M=16$ or $N/M=4$, and the measurement matrix is a filter of length $K=284$. The number of coefficients $L$ is a fixed ratio of $N$ such that the sparsity is $B/M\approx 0.79$.   \label{fig:cpu}}
\end{center}
\end{figure}

The least squares problem in Algorithm~\ref{alg:qomomp} is solved using an iterative method for the normal equation. The relative tolerances are fixed, except for the last level where we decrease the tolerance for higher accuracy. Numerical verification of the computational cost of QOMOMP, Figure~\ref{fig:cpu}, confirms that it scales linearly with the signal size $N$ for typical parameters. Intermediate and final tolerances are set to $\epsilon_i=2\times 10^{-2}$ and $\epsilon_f=3.3\times 10^{-6}$, respectively.

Now, let us return to the energy spectrum estimation problem stated in \eqref{eq:ataproblem}.  Let $L^0$ be the initial sequence of the number of non-zero coefficients at each level for QOMOMP and let the index set $\mathcal J$ specify those levels for which we want to optimize the sequence $L$. With $\app u$ an estimate provided by QOMOMP we iteratively approximate
\begin{equation} \label{eq:minlj}
\min_{L_j} \| \log( E_{A^T A u}) - \log( E_{A^T A \app u}) \|_{w_j}, \quad \forall j \in \mathcal J
\end{equation}
where the weights $w_j$ are constant with support in the range $2^{j-1}<k\leq 2^j$. We put the constraints $L_j\leq L_{j-1}$ in 1-D and $L_j\leq 2L_{j-1}$ in 2-D for $j \in \mathcal J$. We call this low dimensional optimization \emph{spectrum estimation by sparse optimization}\footnote{The code for SpESO with QOMOMP is available at \href{http://github.com}{github.com} as SpESO.} (SpESO). Since the computation of $\app u$ is expensive and the optimization function is non-smooth, we do not solve \eqref{eq:minlj} exactly. Instead, we search amongst values uniformly distributed on a log scale and narrow the search after each iteration. From the linear dependency of QOMOMP on $N$ and the implementation of SpESO, we estimate the overall computational complexity of SpESO to be $\mathcal O(N|\mathcal J|)$.

Our experiments show that decoupling the matrix used in SpESO from the one used in QOMOMP improves the convergence properties. By that, we mean that the measurement matrix is split horizontally into two parts $\prim A$ and $\seco A$, giving a set of measurements $\prim g=\prim Au$ and $\seco g=\seco Au$. For QOMOMP we use $\prim \Psi=\prim A\Phi$ and $\prim g$ and for SpESO we use $\seco A^T \seco A$, and {\em vice versa\/}. The two estimated spectra are then combined proportionally to their relative errors. A simplistic argument for the decoupling is that since QOMOMP minimizes the error $A\app u-g$ to a small or zero value regardless of $L$, then the difference between $A^T A u$ and $A^T A \app u$ will be small and \eqref{eq:minlj} will not converge to any meaningful minimum. By using two separate matrices this problem disappears and results in a better correlation between a good choice of $L$ and a low energy spectrum error. The downside is that QOMOMP only uses half of the measurements for each estimation.

\section{Analysis of the performance of SpESO for ideal signals\label{sec:proof}}
We now analyze mathematically the convergence and accuracy of SpESO. Let us consider the restricted isometry property (RIP) of the CS matrices that  determines the accuracy of reconstructions. The restricted isometry constant  of a matrix $\Psi$ is the smallest number $\delta_B$ such that \cite{candes2005decode,candes2008resiso}
\begin{equation}\label{eq:ripdef}
(1-\delta_B)\|x\| \leq \|\Psi x\| \leq (1+\delta_B)\|x\|
\end{equation}
holds for all $x$ at most $B$-sparse. If the OMP algorithm is applied with a matrix $\Psi$ satisfying $\delta_{B+1}<1/(\sqrt B +1)$, then it recovers a $B$-sparse signal exactly \cite{wang2012omplimit}. The proof is mainly concerned with showing that at each iteration the index chosen is in the true support $T$. Given the true support at the final iteration, the reconstruction is trivial.

Assume $T$ is the true support of the best $B$-term approximation $u_{B}$. In the case of a perfect oracle where $\Omega=T$ in QOMOMP, the solution to the final least squares problem is
\begin{equation}
\tapp u_T = \Psi_T^{+} \Psi \hat u = \Psi_T^{+} (\Psi_T \hat u_T + \Psi_{T^c} \hat u_{T^c}),
\end{equation}
where $\Psi_T^{+}$ is a pseudo-inverse. With $\Psi_T^{*}\Psi_T$ non-singular ($\delta_B<1$) we get $\tapp u_T = \hat u_T + \Psi_T^{+} \Psi_{T^c} \hat u_{T^c}$. Therefore, the error is
\begin{equation}
\|\hat u_T - \tapp u_T \| = \| \Psi_T^{+} \Psi_{T^c} \hat u_{T^c} \|
\end{equation}
or, with $\Phi$ orthonormal 
\begin{equation}
\| u - \app u \|^2 = \|\hat u_{T^c}\|^2 + \| \Psi_T^{+} \Psi_{T^c} \hat u_{T^c} \|^2 \leq \| \hat u-\hat u_B \|^2 + \frac{1}{1-\delta_B}\| \Psi(\hat u-\hat u_B) \|^2
\end{equation}
(the inequality follows from RIP \cite{needell2008cosamp}). For a compressible signal $u$, the error depends on the best $B$-term approximation error $\|\hat u_{T^c}\|=\| u- u_B\|$ and the least squares error term, which depends on the RIP of the matrix $\Psi$. Given $u$ is $B$-sparse ($u=u_B$), the error vanishes.

Now consider our QOMOMP method in a very simple 1-D setting to obtain some quantitative performance estimates. Let QOMOMP be applied to a signal $u$ with a power law energy spectrum $k^{-\alpha}$, where $0<\alpha\le 2n+1$ is limited by the number of vanishing moments $n$ of the wavelet used in the sparsity system. The variance of the wavelet coefficients at each level then scales like $\Var (\hat d_i^j)_i \sim 2^{-j\alpha}$~\cite{Perrier/etal:1995}. Assuming $u$ is a Fourier synthetic signal like those considered in section~\ref{sec:num}, then $\hat d_i^j$ for each level is well approximated as  i.i.d.\ with a Gaussian distribution and zero mean. If $\Omega_{J_0}=\cup_{j<J_0} \Gamma_j$ is the initial QOMOMP setup, the probability that the true support $T_B$ of the best $B$-term approximation contains  $\Omega_{J_0}$ is
\begin{equation}
P_{J_0}=\Pr (\Omega_{J_0} \subset T_B)= \prod_{j<J_0} \Pr (|\hat d_i^j|\geq \epsilon_B)^{2^j},
\end{equation}
where the threshold $\epsilon_B$ is the best $B$-term threshold such that
\begin{equation}
B/N = \frac{1}{N}\sum_{j<J}\sum_{i} \Pr (|\hat d_i^j|\geq \epsilon_B)=\frac{1}{N}\sum_{j<J} 2^j \left[ 2-2F(\epsilon_B 2^{j\alpha /2}) \right]
\end{equation}
where $F$ is the standard cumulative distribution.  

Assume now, as in the 1-D experiments in section~\ref{sec:num}, that $N=2^{15}$ and $B=N/8$. If $J_0=5$ then $P_{5}=82\%$ for energy spectrum slope $\alpha=5/3$ and $P_{5}=99.5\%$ for energy spectrum slope $\alpha=3$. There is a reasonable probability that {\em every\/} coefficient in $\Omega_{J_0}$ is in the true support. We can also consider how many coefficients $L_j$ should be estimated at finer levels $j\ge J_0$. At the single level $j=7$  94.9\% of the coefficients are included in $u_B$ for $\alpha=5/3$ and 99.4\% are included for $\alpha=3$. Thus, the exact value for the number of coefficients to estimate at this level is $L_7 = 0.95\times 2^7$ for $\alpha=5/3$.  Recall that in practice the sequence $\{L_j\}$ must be estimated {\em a priori\/}, but this sort of analysis gives us a good {\em ansatz} for determining it. Note that adding some excess coefficients that are not in the true support is not a serious problem; the algorithm in \cite{wang2012gomp} defines an iteration as successful if at least one coefficient is correctly chosen. 

A rigorous analysis of QOMOMP would involve the RIP of $\Psi$ in addition to the distribution of the wavelet coefficients. However, since the CS algorithm estimates are usually conservative, they are not a good indicator the actual performance of the method.  Therefore, in the next section we rely instead on a wide range of representative computational experiments to assess the actual performance of SpESO.

\section{Numerical tests of SpESO\label{sec:num}} 
\subsection{Turbulence test signals and computational parameters}
To test the effectiveness of SpESO we need signals with energy spectra and arbitrary power law scaling. For this purpose, synthetic turbulence signals with power law energy spectra are particularly appropriate, in addition to experimental signals and data from numerical simulations of turbulence.  

In our results, synthetic signal type $(x,y)$ denotes a signal with two energy spectrum power laws $-x$ and $-y$, i.e., $E(k)\sim k^{-x}$ and $E(k)\sim k^{-y}$, split at $k=N/32$ in 1-D (unless specified otherwise) and $k=N/8$ in 2-D.  Signal $(x)$ denotes a signal with a single power law. $F(x,y)$ denotes a synthetic Fourier signal, and $W(x,y)$ denotes a synthetic wavelet signal.  The Fourier and Wavelet synthetic signals are described below. Note that signals with a change in slope are particularly challenging for energy spectrum estimation when this change occurs at wavenumbers larger than the Nyquist wavenumber corresponding to the Shannon sampling rate since the second slope would be not be resolvable using a standing Shannon sampling technique.

\begin{figure}[tbp]
\centering\footnotesize
\renewcommand{\tabcolsep}{2pt}     % width
\begin{tabular}{ccc}
Fourier (linear $\zeta_p$) &  Wavelet (concave $\zeta_p$) & Hot-Wire \\
\includegraphics[width=\figwidthes]{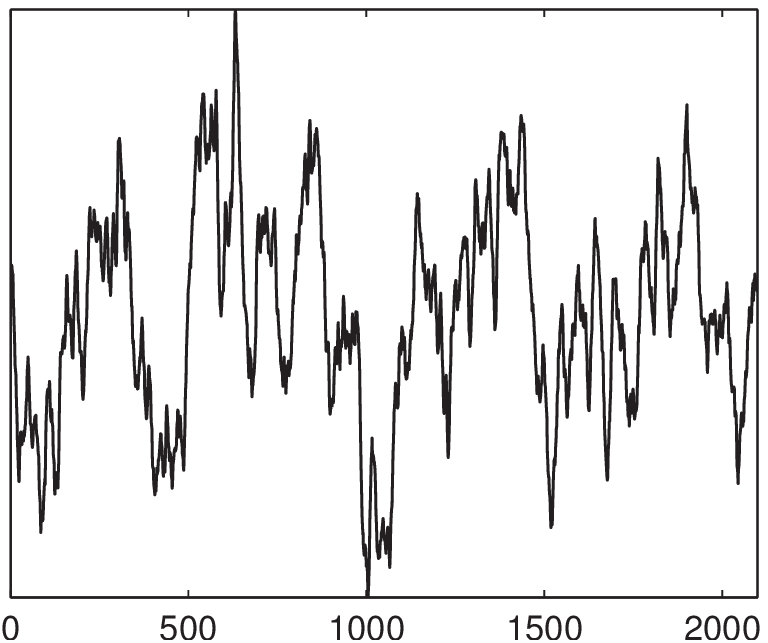}&
\includegraphics[width=\figwidthes]{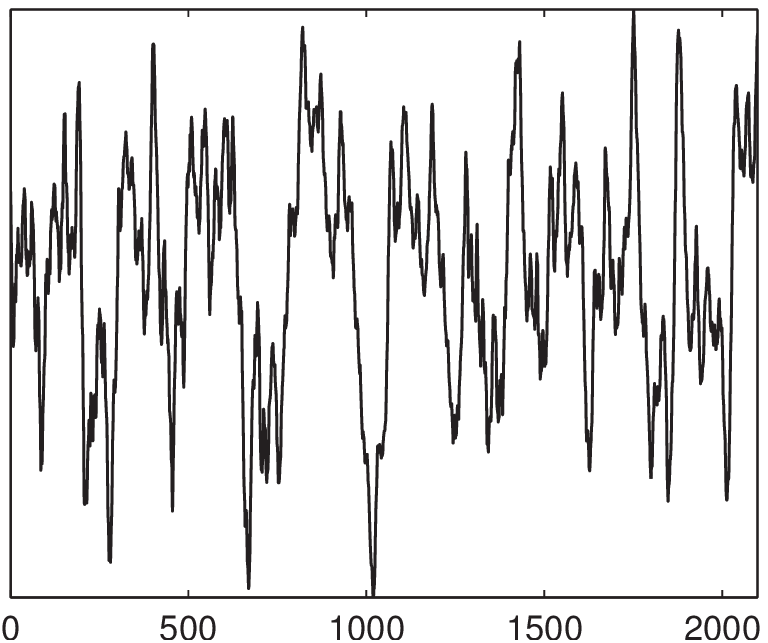}&
\includegraphics[width=\figwidthes]{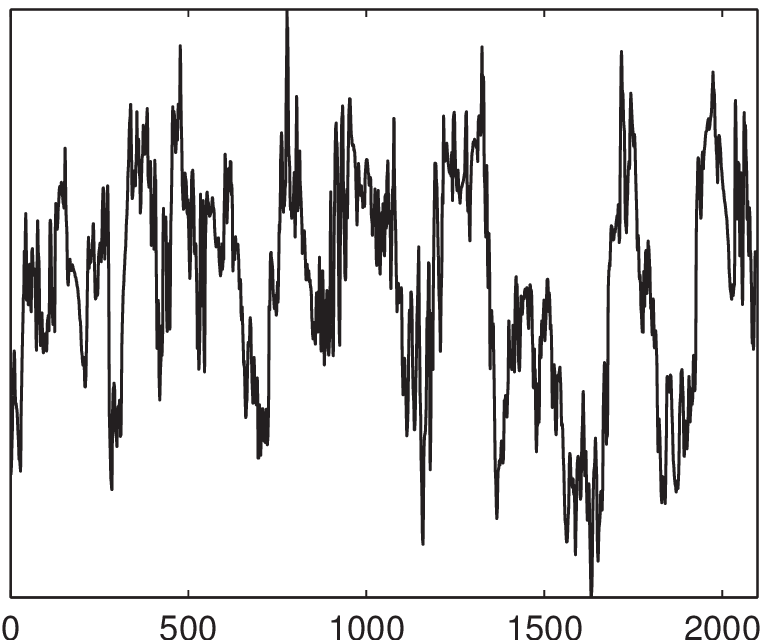} \\
Fourier (linear $\zeta_p$) &  Wavelet (concave $\zeta_p$) & DNS (Johns Hopkins) \\
\includegraphics[width=\figwidthes]{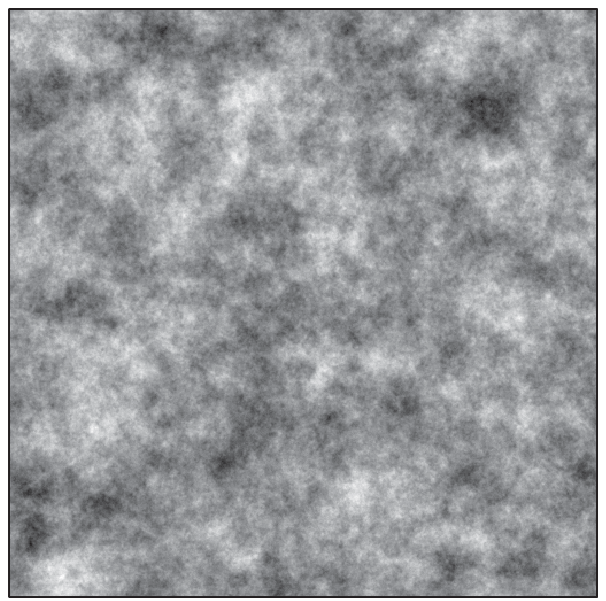}&
\includegraphics[width=\figwidthes]{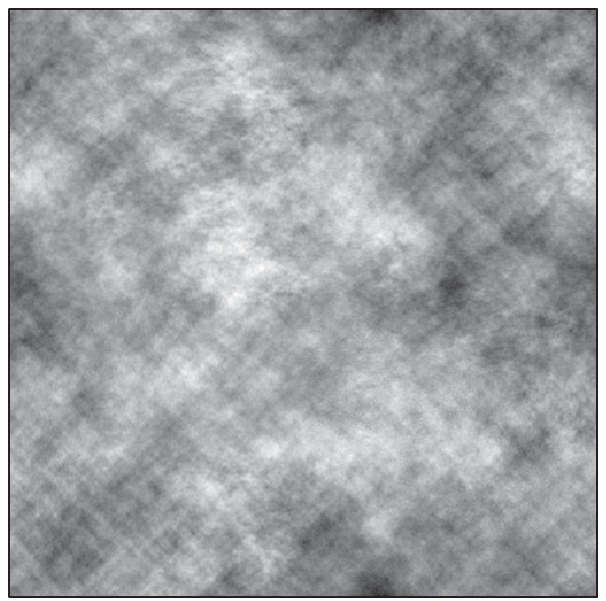}&
\includegraphics[width=\figwidthes]{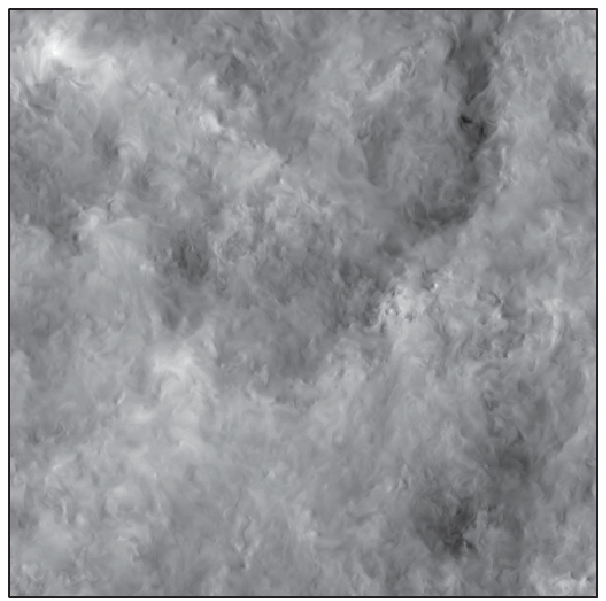}
\end{tabular}
\caption{1-D and 2-D turbulence test signals. The Fourier and wavelet synthetic signals both have $(5/3,3)$ energy spectra typical of a 2-D flow, while the experimental Hot-Wire signal and the DNS data have a signal power-law scaling $(5/3)$ typical of 3-D flow.  Note that the Fourier signal is non-intermittent (with linear $\zeta_p$), while the wavelet signal has been designed to have more realistic intermittent statistics (with concave $\zeta_p$). \label{fig:ks}}
\end{figure}

The first method constructs a synthetic signal in Fourier space. The Fourier coefficients of the signal $u$ are determined by the desired energy spectrum but with random complex phases,
\begin{equation}
\hat u_k = \sqrt{E(k)} e^{i \theta_k},
\end{equation}
where $\theta_k$ are i.i.d.\ uniformly on $[0,2 \pi)$ and $E(k)$ is the specified spectrum. In higher dimensions, the coefficients $\hat u[\mathbf k]$ for which $|\mathbf k|=k$ have variance proportional to $E(k)$. The resulting signal is  homogeneous  Gaussian statistics.  Typical realizations of the Fourier-based synthetic 1-D test signal (top left) and 2-D data (bottom left) are shown in Figure~\ref{fig:ks}.

In addition to the random phase Fourier synthetic signal described above, we also consider a synthetic multiscale signal, generated by  a random process in wavelet space. This allows use to generate a synthetic signal that is closer to a true turbulent signal. Kolmogorov's original statistical theory of turbulence~\cite{frisch1999turbulence} predicts a structure function scaling $S_p(r)  = \frac{1}{L}\int_{0}^{L} |u(x)-u(x+r)|^{p} \text{d}x \sim r^{\zeta_p}$ with $\zeta_p=p/3$, in the limit of infinite inertial subrange, and this is approximately the scaling produced by the Fourier-based synthetic signal process described above.  However, actual experimental measurements show that as a result of intermittency $\zeta_p$ increases more slowly than linearly with $p$, i.e. it is concave.  The lack of intermittency in the Fourier synthetic signal means that the resulting data sequence is more homogeneous locally than real turbulence.  In order to assess the ability of SpESO to cope with intermittency we have also use the wavelet-based method of~\cite{benzi1993affine} to synthesize a signal with a more realistic concave function $\zeta_p$. The wavelet-based signal has a realistic concave, intermittent scaling of structure function exponents $\zeta_p$ while the Fourier signal has a non-physical slightly convex scaling. The scaling of the energy spectrum is then  defined implicitly by the second-order structure function.  Typical realizations of the Wavelet synthetic 1-D test signal  $W(5/3,3)$ (top middle) and  2-D data (bottom middle) are shown in Figure~\ref{fig:ks}.  Note that we generate this wavelet-based synthetic data using symmlet~12 wavelets with six vanishing moments, rather than the Coiflet wavelets used for the energy spectrum estimation algorithm, to ensure that the data is independent of the sparsity system used in the compressive sampling.

Finally, we consider two realistic turbulence data sets: a 1-D times series measurement of a single velocity component of an axisymmetric jet~\cite{nobach1996a}, and a 2-D slice of a 3-D direct numerical simulation (DNS) of homogeneous isotropic turbulence~\cite{li2008jhdata,perlman2007jhdata}.   The 1-D data is from hot-wire measurements at 20kHz and $\text{Re}=4\times10^4$ and a typical section is shown in Figure~\ref{fig:ks} (top right).  Note that the energy spectrum of this data has a signal power law scaling $k^{-5/3}$.  The second data set is from a high Reynolds number 3-D pseudo-spectral turbulence simulation stored in the Johns Hopkins University (JHU) turbulence database cluster~\cite{li2008jhdata,perlman2007jhdata}.  The Taylor scale Reynolds number of this flow is $ R_\lambda\sim 433$ (corresponding to $\text{Re}\approx 2\times 10^5$) . This simulation has a resolution of $1024^3$, and therefore the 2-D slice has a resolution of $1024^2$.  To simplify the analysis, we consider a single velocity component.  A typical example of this data is shown Figure~\ref{fig:ks} (bottom right).  

The results are computed using the Coiflet~18 wavelet basis with six vanishing moments for 1-D signals and Coiflet~12 wavelet basis with four vanishing moments for 2-D signals. The random filter is i.i.d.\ uniform in $\{\pm1\}$ and the length for all 1-D cases is $K=284$. For the tree function, the threshold function is $\Lambda : a\to \frac{1}{2}\sqrt{\Var(a)}$, and $\beta=3$ ($\beta=2$ in 2-D). The set of levels to optimize over is approximately $\mathcal J=\{j : j>\log_2 (M/2)\}$. The {\em a priori\/}-based initial guess of coefficients $L^0$ is set approximately to the number of coefficients of each level in a typical best $M$-term approximation and the level $J_0=5$ in 1-D and $J_0=4$ in 2-D.  Logarithmic scale averages of spectra are essentially geometric means of spectra. 

Dashed lines in the figures show the theoretical slope of the power law parts of the spectrum. In each case, we compare SpESO with the original signal, best $M$-term wavelet approximation and the usual fixed rate Shannon sampling.  In many cases we also compare results with the best $M/2$-term approximation and the LOMP method. 

The differences between each of the spectrum estimation methods are as follows. The $M$-best term approximation  first takes the wavelet transform of the entire signal and then selects the largest $M$ wavelet coefficients for the estimation. It is therefore not a sampling scheme, but rather an ideal benchmark to which the CS methods are compared.  We expect that the $M$-best term approximation to be the best possible estimate using $M$ samples.  The Shannon scheme subsamples $u$ at uniform rate (without low-pass filtering), followed by a Fourier interpolation. LOMP is an iterative CS method like the well-known OMP that estimates a few ($L_0\geq 1$) coefficients at a time, without using any {\em a priori\/} information or the tree structure of the wavelet coefficients. Finally, SpESO is a CS optimization method that uses {\em a priori\/} information to minimize the logarithmic scale error of the energy spectra.  It also enforces a realistic tree structure for the estimated wavelet coefficients.

\subsection{Results for 1-D Signals\label{sec:1-Dcases}} 
\begin{figure}[tbp]
\centering\footnotesize
\renewcommand{\tabcolsep}{2pt}     % width
\begin{tabular}{ccc}
\rotatebox{90}{ \qquad\qquad\qquad Fourier } &
\includegraphics[width=\figwidtht]{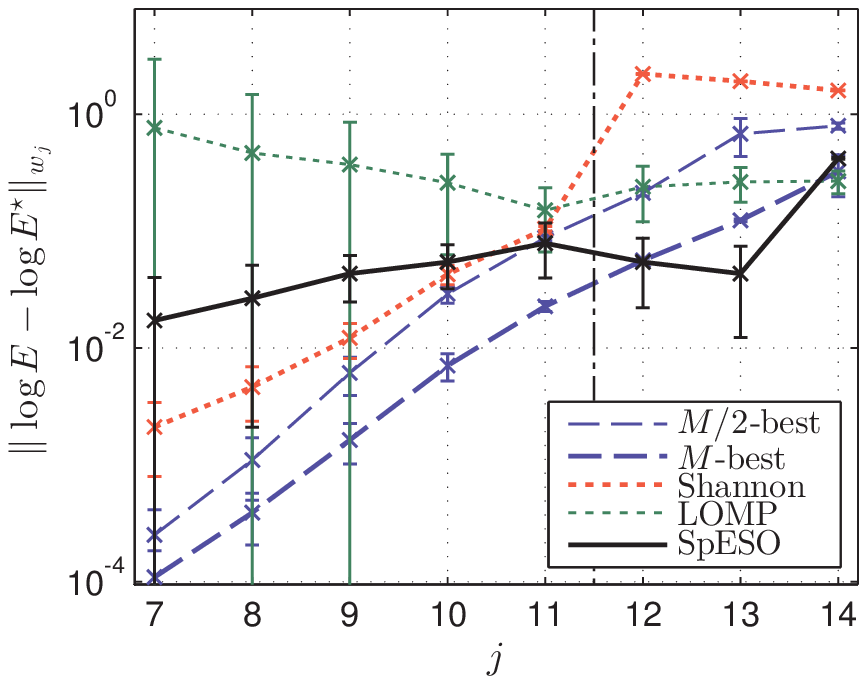}&
\includegraphics[width=\figwidtht]{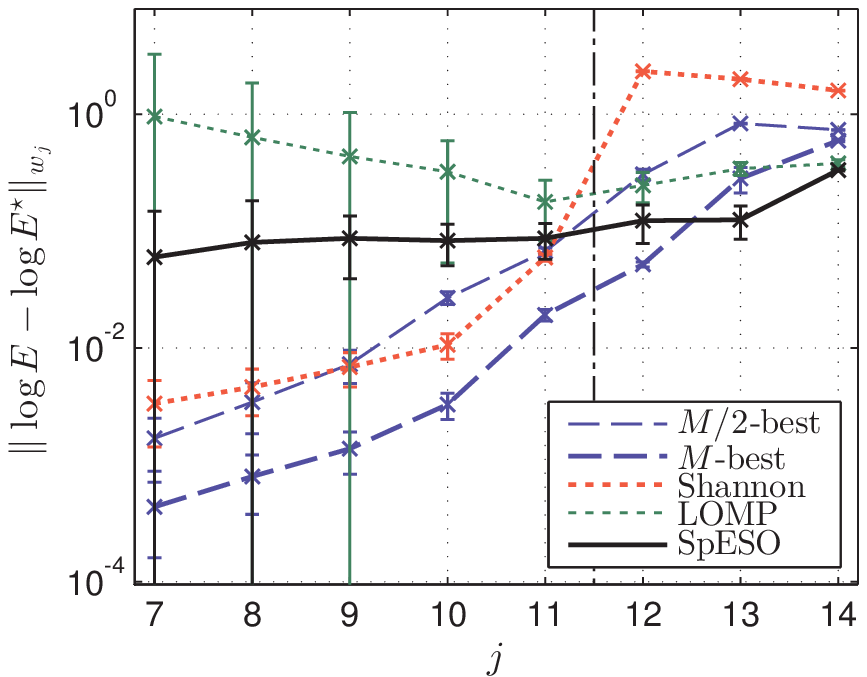}\\ 
\rotatebox{90}{ \qquad\qquad\qquad Wavelet } &
\includegraphics[width=\figwidtht]{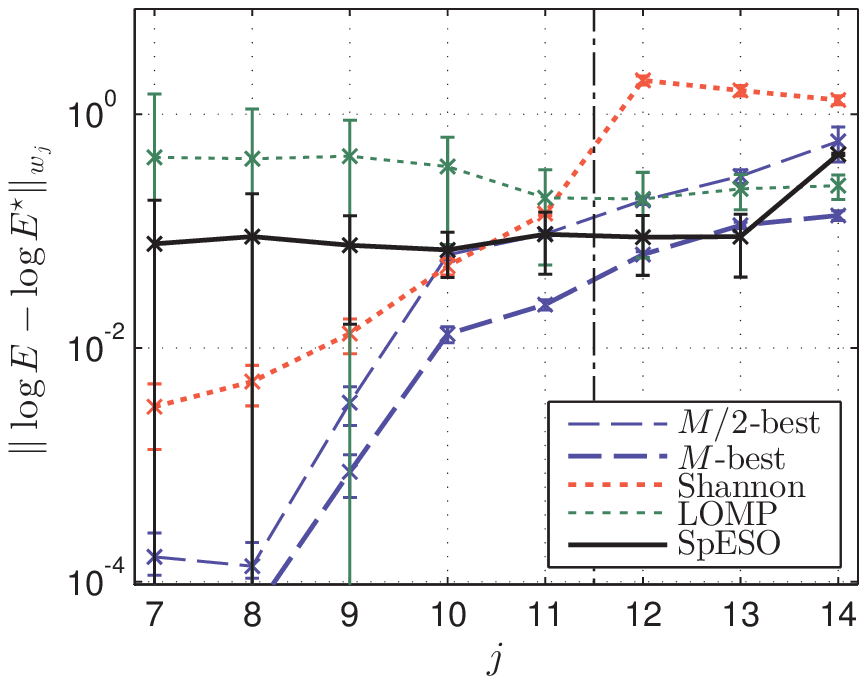}&
\includegraphics[width=\figwidtht]{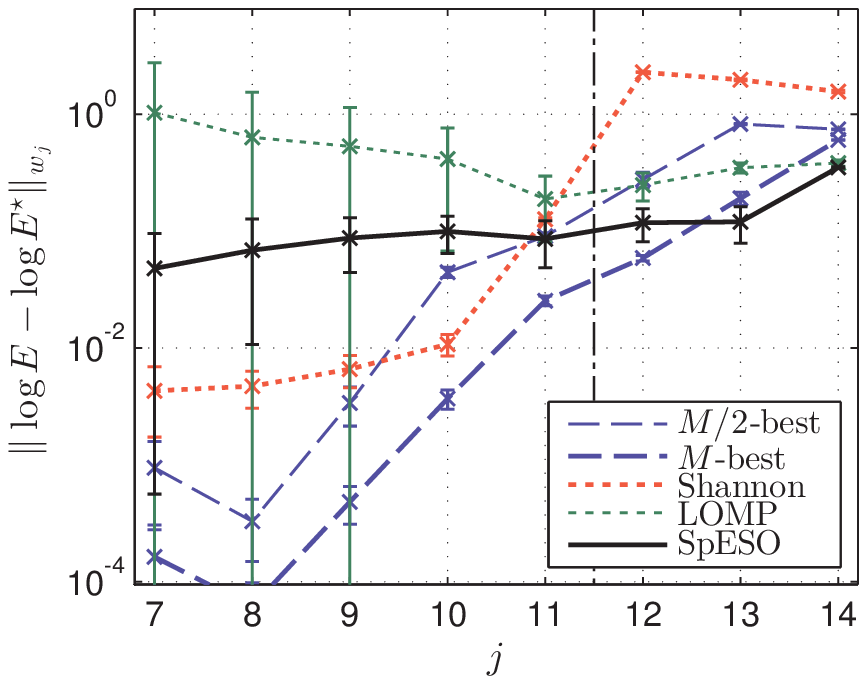}\\ 
 &  $(3,5/3)$ &  $(5/3,3)$
\end{tabular}
\caption{Logarithmic scale errors of spectrum estimations versus the weight location $j$, on a logarithmic scale, with $N=2^{15}$ and $N/M=8$ (Nyquist wavenumber at dash-dot vertical line). The norm weights $w_j$ are compact around wavenumbers corresponding to scale $j$ and $\sum_k w_j(k)=1$. The plots show mean curves and standard deviation bars of 64 random simulations. The signals are 1-D with spectrum slopes splitting at $k=N/32$ ($j=10.5$). Fewer samples at large scales (small $j$) result in larger error bars.\label{fig:logerror}}
\end{figure}

\begin{figure}[tbp]
\centering\footnotesize
\renewcommand{\tabcolsep}{2pt}     % width
\begin{tabular}{ccc}
\rotatebox{90}{ \qquad\qquad\qquad Fourier } &
\includegraphics[width=\figwidtht]{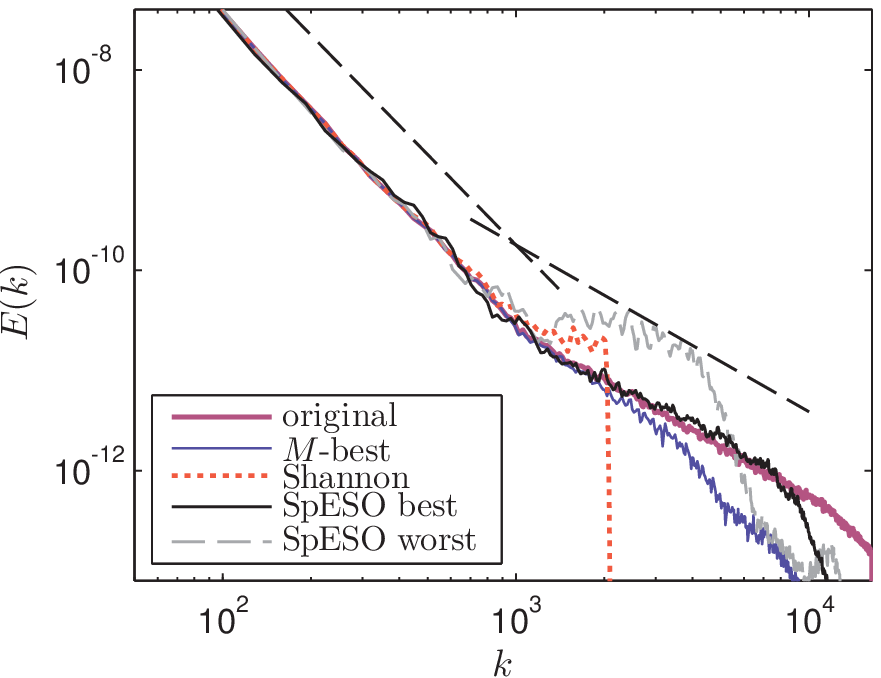}&
\includegraphics[width=\figwidtht]{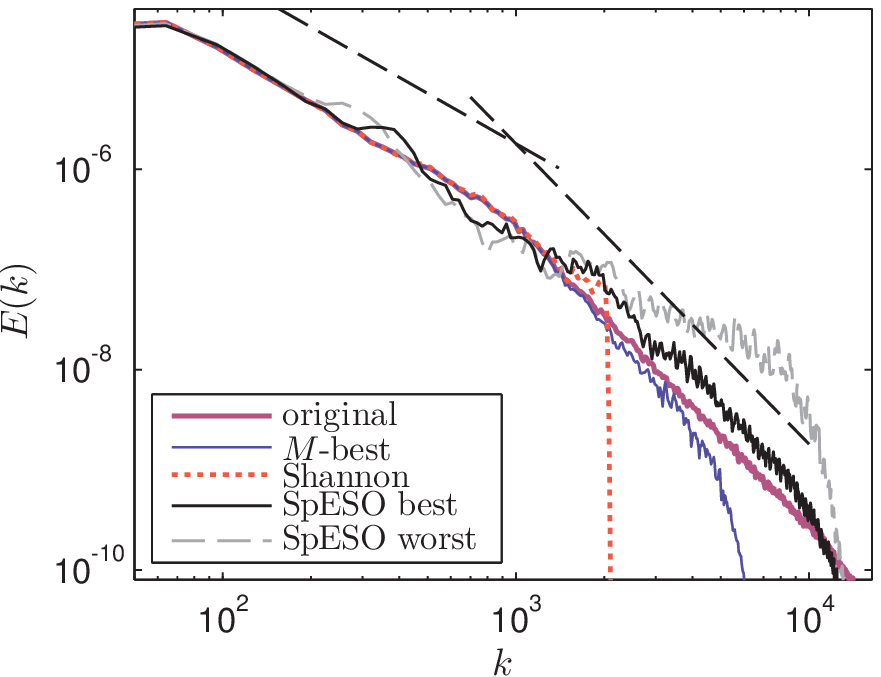}\\ 
\rotatebox{90}{ \qquad\qquad\qquad Wavelet } &
\includegraphics[width=\figwidtht]{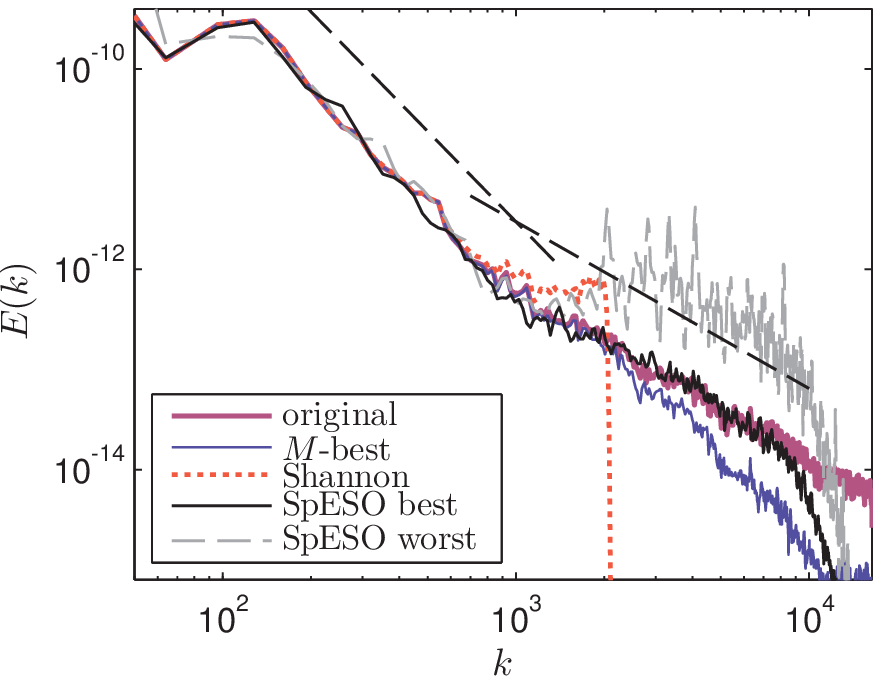}&
\includegraphics[width=\figwidtht]{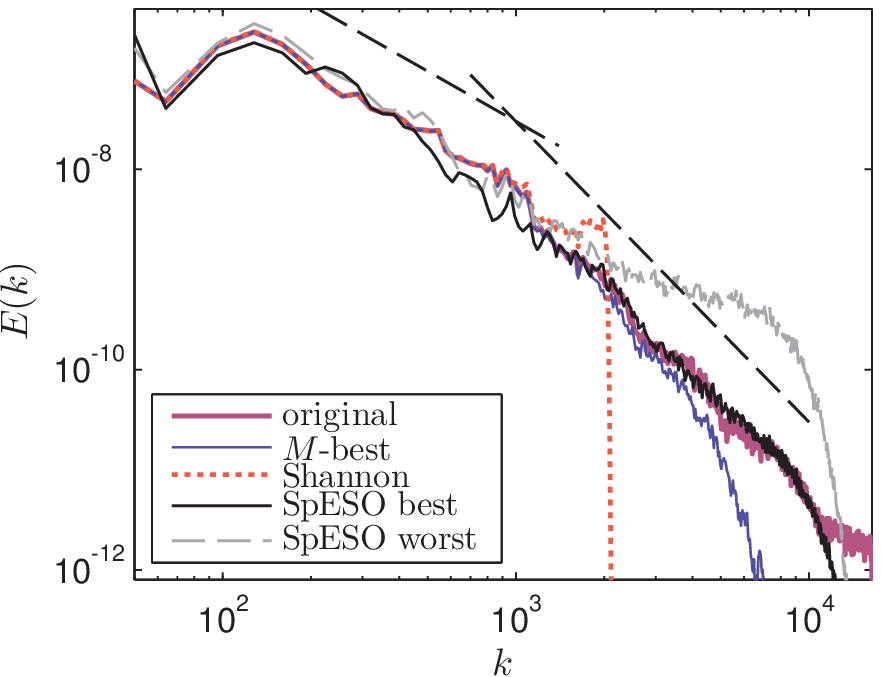}\\ 
 &  $(3,5/3)$ &  $(5/3,3)$
\end{tabular}
\caption{Representative energy spectrum estimations of 1-D signals, with $N=2^{15}$ and $N/M=8$, out of the 16 simulations. The slope of the energy spectrum changes at $k=N/32=1024$, smaller than the Nyquist wavenumber $k=N/16=2048$. The best and worst cases of SpESO reveal the variation of its approximations.  \label{fig:logcases}}
\end{figure}

\begin{figure}[tbp]
\centering\footnotesize
\renewcommand{\tabcolsep}{2pt}     % width
\begin{tabular}{ccc}
\rotatebox{90}{ \qquad\qquad\qquad Fourier $(5/3,3)$ } &
\includegraphics[width=\figwidtht]{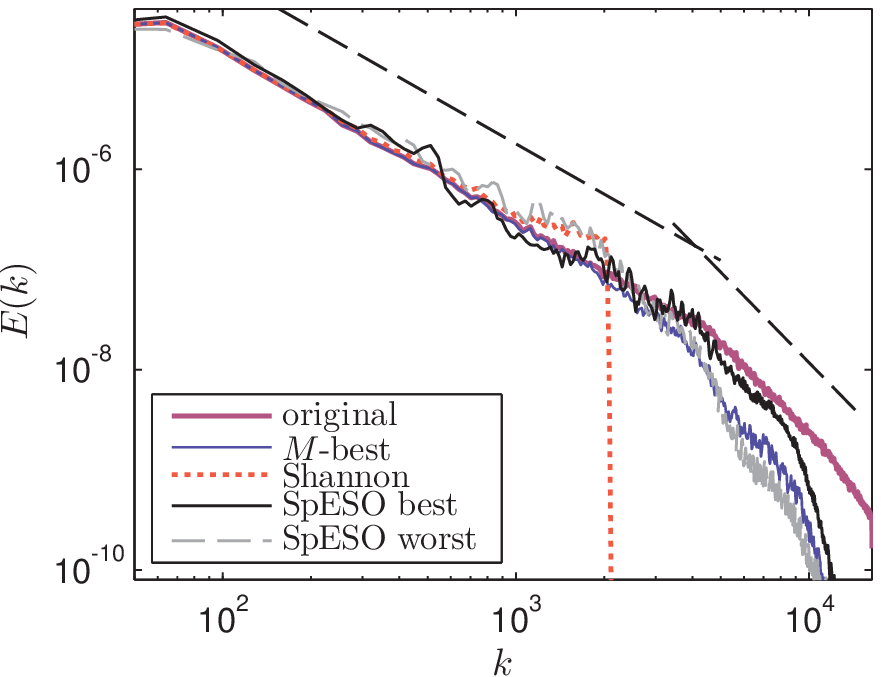}&
\includegraphics[width=\figwidtht]{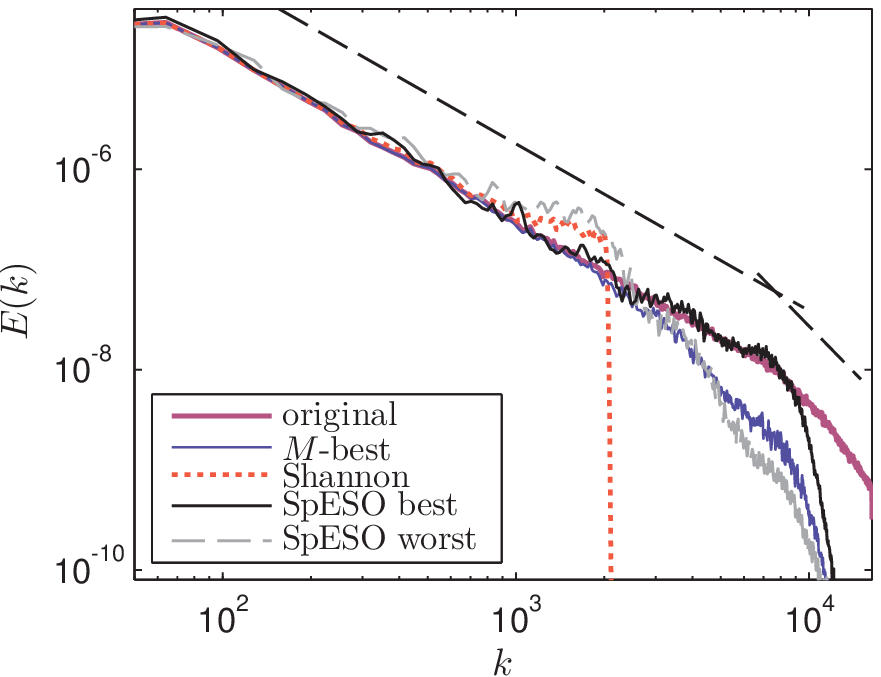} \\
& $k=N/8$ &  $k=N/4$
\end{tabular}
\caption{Representative energy spectrum estimations similar to Figure~\ref{fig:logcases} (upper right), but with the slope changing at $k=N/8=4096$ (left) and $k=N/4=8192$ (right). \label{fig:logcases2}}
\end{figure}

\begin{figure}[tbp]
\centering\footnotesize
\renewcommand{\tabcolsep}{2pt}     % width
\begin{tabular}{cccc}
\rotatebox{90}{ \quad Wavelet $(3,5/3)$ } &
\includegraphics[width=\figwidthe]{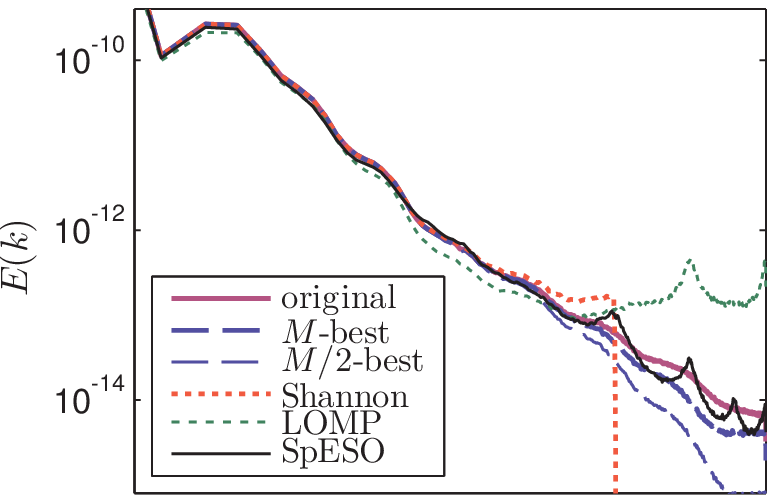}&
\includegraphics[width=\figwidthes]{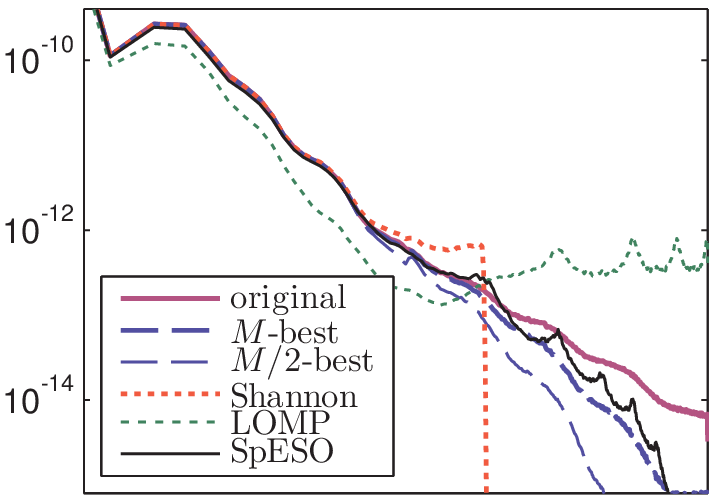}&
\includegraphics[width=\figwidthes]{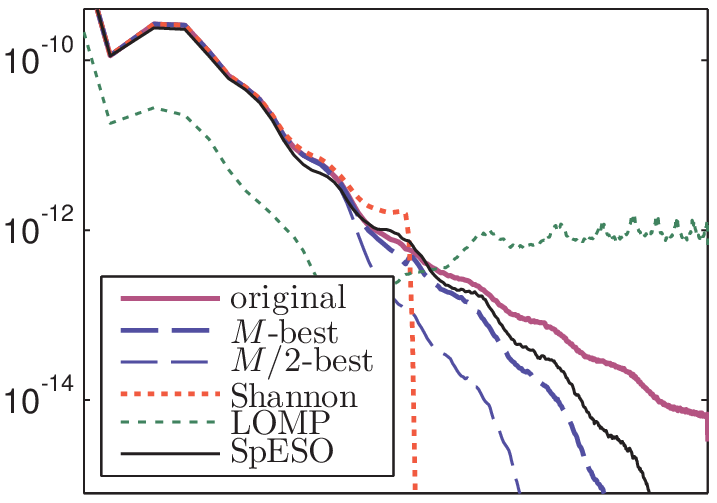}\\ 
\rotatebox{90}{ \quad Fourier $(3,5/3)$ } &
\includegraphics[width=\figwidthe]{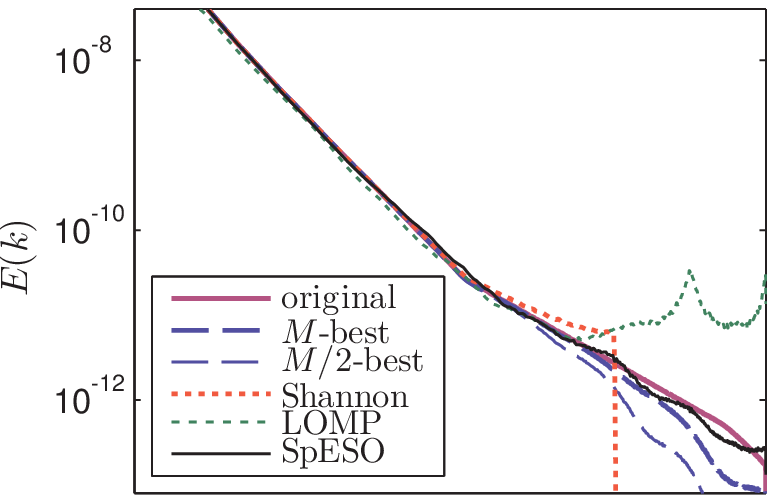}&
\includegraphics[width=\figwidthes]{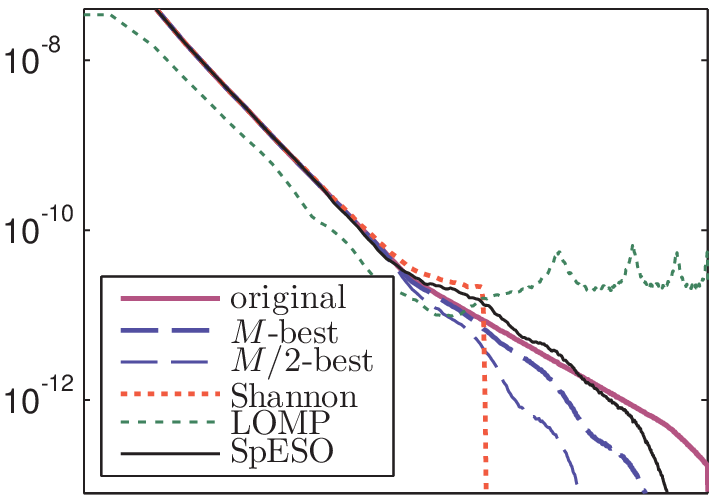}&
\includegraphics[width=\figwidthes]{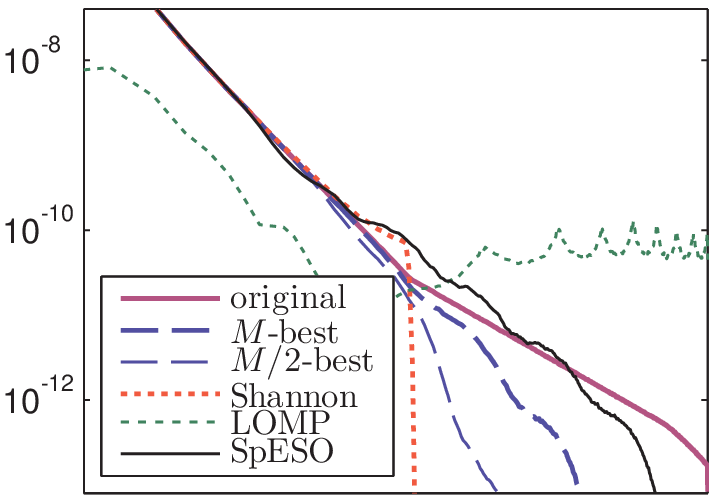}\\ 
\rotatebox{90}{ \quad Wavelet $(5/3,3)$ } &
\includegraphics[width=\figwidthe]{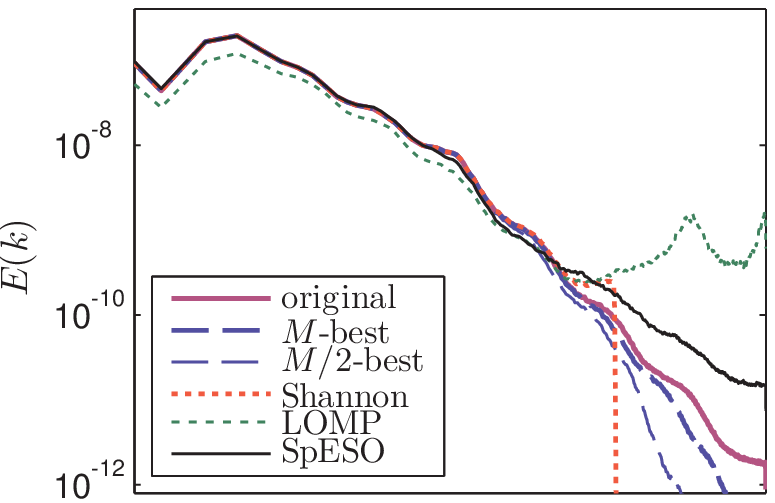}&
\includegraphics[width=\figwidthes]{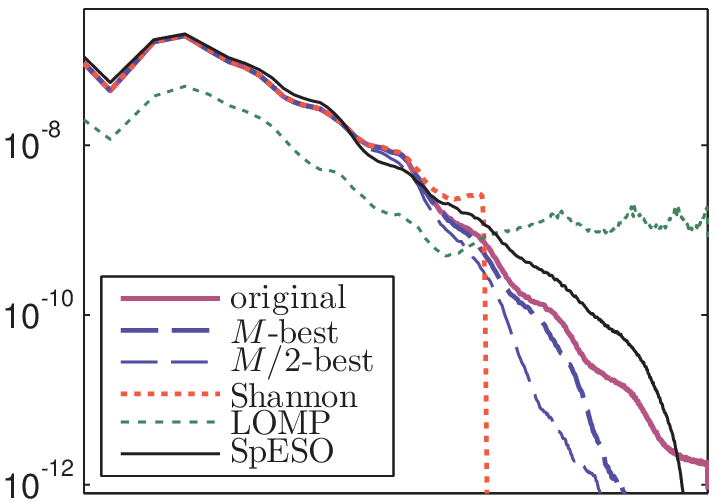}&
\includegraphics[width=\figwidthes]{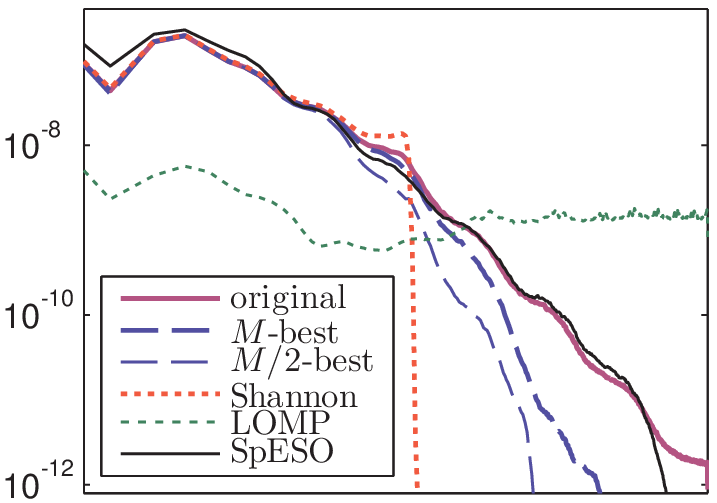}\\ 
\rotatebox{90}{ \qquad Fourier $(5/3,3)$ } &
\includegraphics[width=\figwidthe]{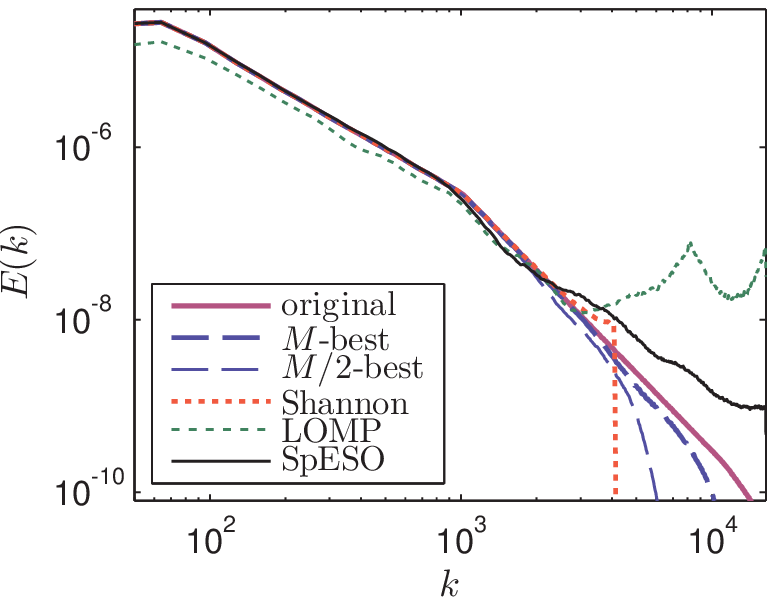}&
\includegraphics[width=\figwidthes]{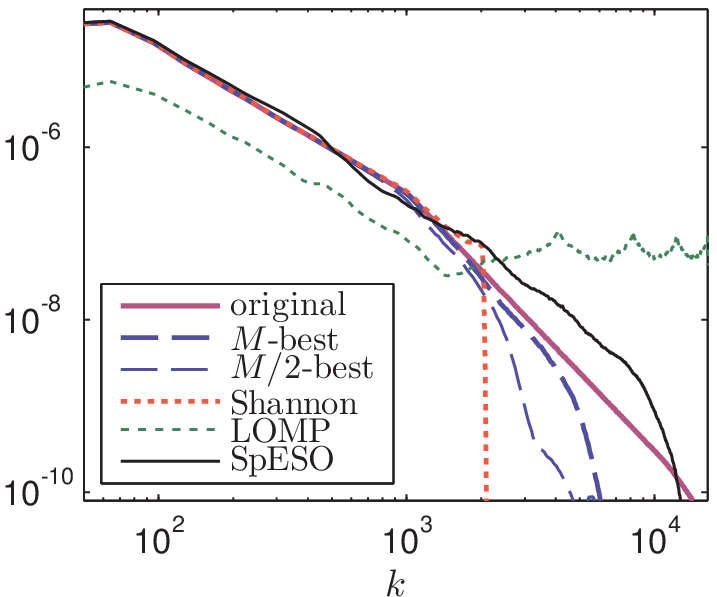}&
\includegraphics[width=\figwidthes]{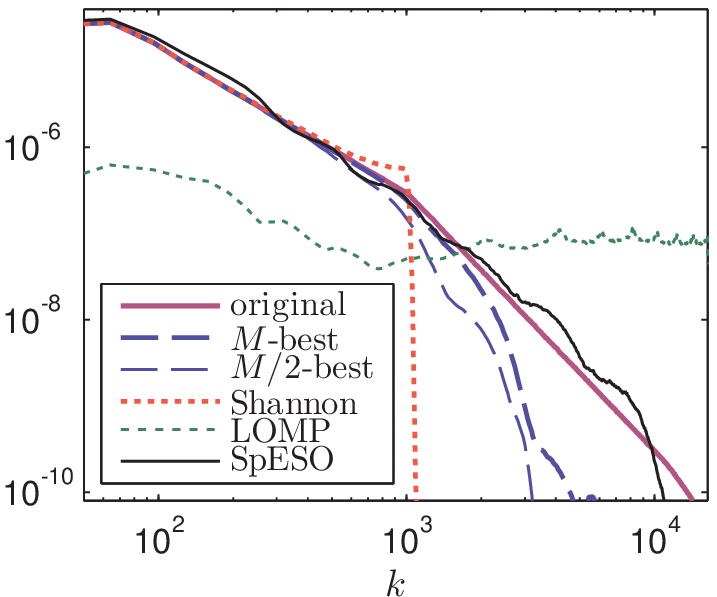}\\ 
 &  $N/M=4$ &  $N/M=8$ &  $N/M=16$  
\end{tabular}
\caption{ Logarithmic scale averages of spectrum estimates for 1-D signals and various sampling ratios. Corresponds to Tables \ref{tab:slopeerrorone} and \ref{tab:slopeerror}.    \label{fig:logspecaver3}}
\end{figure}

\begin{figure}[tbp]
\centering\footnotesize
\renewcommand{\tabcolsep}{2pt}     % width
\begin{tabular}{ccc}
%\rotatebox{90}{ \qquad\qquad\qquad Fourier } &
%\includegraphics[width=\figwidtht]{./a1_32a}&
%\includegraphics[width=\figwidtht]{./a1_32b}\\ 
\rotatebox{90}{ \qquad\qquad\qquad Wavelet } &
\includegraphics[width=\figwidtht]{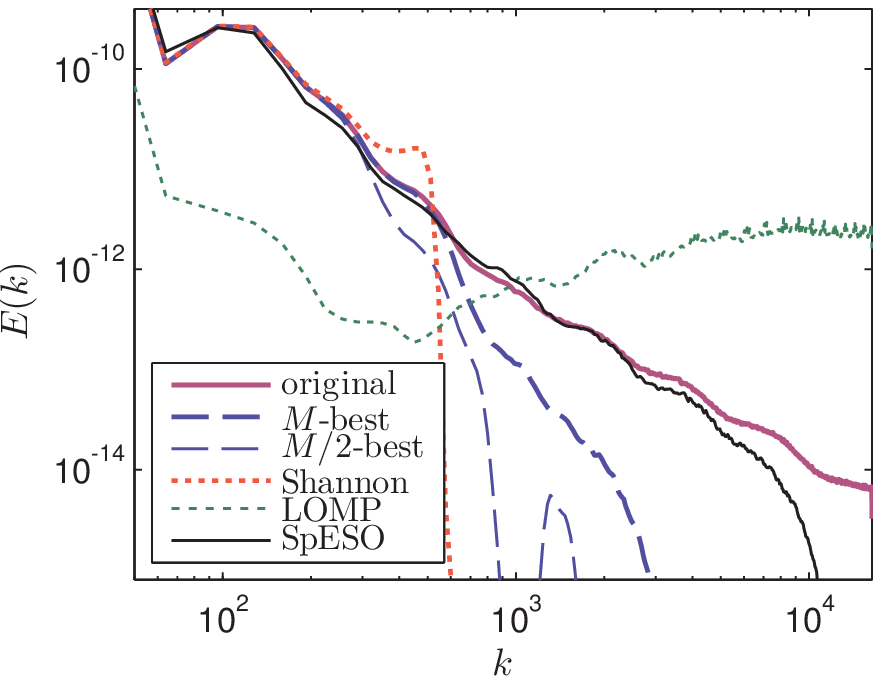}&
\includegraphics[width=\figwidtht]{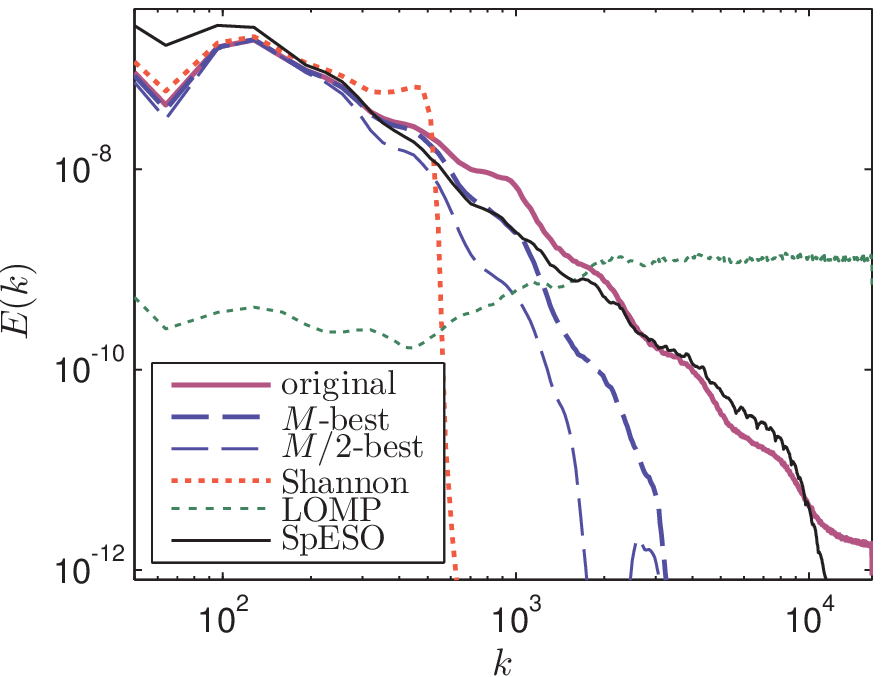}\\ 
 &  $(3,5/3)$ &  $(5/3,3)$
\end{tabular}
\caption{ Logarithmic scale averages of spectrum estimates for 1-D signals for a high sampling ratio where the slope of the energy spectrum changes at a wavenumber higher than the Nyquist wavenumber. The signal lengths are $N=2^{15}$, the sampling ratio is $N/M=32$, and the number of simulations is 64. The slope changes at $k=N/32=1024$ which is  larger than the Nyquist sampling wavenumber $k=512$ corresponding to the sampling ratio 32. \label{fig:logspecaver32} }
\end{figure}

\begin{figure}[tbp]
\centering\footnotesize
\renewcommand{\tabcolsep}{2pt}     % width
\begin{tabular}{cccc}
\rotatebox{90}{ \qquad\quad Hot-wire } &
\includegraphics[width=\figwidthe]{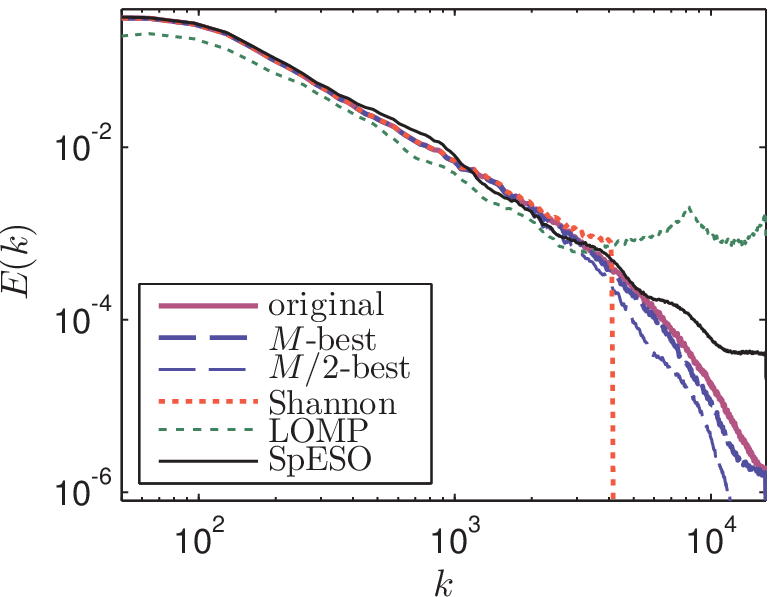}&
\includegraphics[width=\figwidthes]{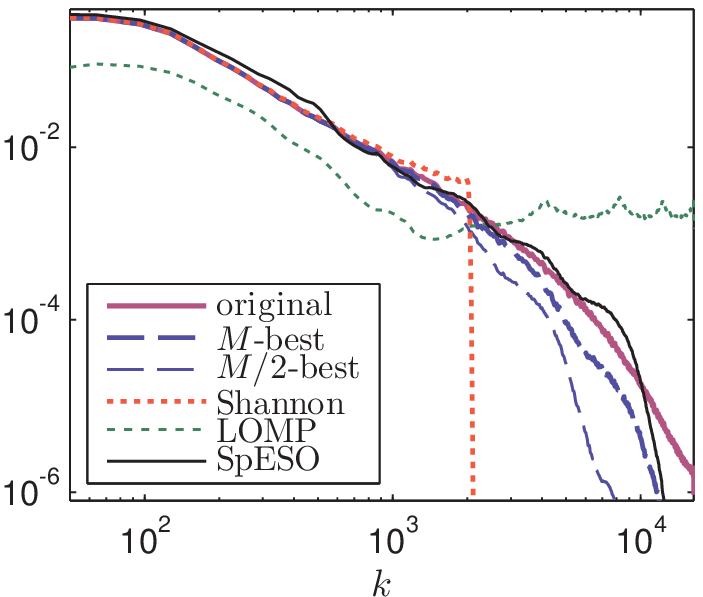}&
\includegraphics[width=\figwidthes]{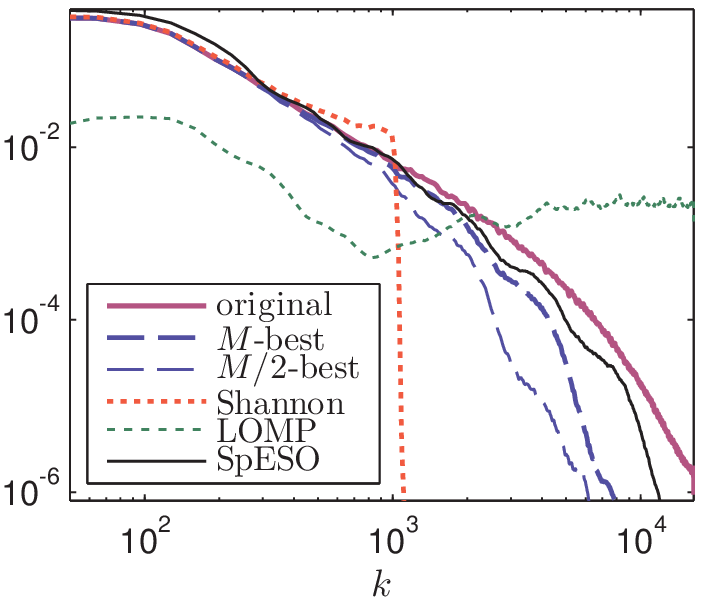}\\ 
& $N/M=4$ & $N/M=8$ & $N/M=16$ 
\end{tabular}
\caption{ Logarithmic scale averages of spectrum estimates for a 1-D hot-wire measurement. The signal length is $N=2^{15}$ and the number of simulations is 64. \label{fig:logspecaverrhh}}
\end{figure}

The performance of SpESO compared to other methods is tested numerically for a range of signal length to measurement length ratios $N/M$ (i.e. sampling ratios). The other methods are fixed rate Shannon sampling, best $M/2$ and $M$-term wavelet approximations in the Coiflet~12 basis, and the CS reconstruction method LOMP.  For the tested $N/M$ ratios 4, 8, 16, 32, SpESO and LOMP have the nearest ratio satisfying \eqref{eq:filterm}, namely 3.97, 7.93, 15.86, 31.72, respectively. The results for the Fourier (F) and Wavelet (W) synthetic 1-D signals are shown in Figures~\ref{fig:logerror}, \ref{fig:logcases}, \ref{fig:logcases2}, \ref{fig:logspecaver3}, \ref{fig:logspecaver32}, and Tables~\ref{tab:slopeerrorone} and \ref{tab:slopeerror}. The results for hot-wire signals are shown in Figure~\ref{fig:logspecaverrhh}.

The energy spectrum errors for each level shown in Figure~\ref{fig:logerror} behave as expected for Shannon sampling: they increase dramatically at the Nyquist wavenumber. The best-term approximations are the most accurate method at large scales, but have a steeply rising error at smaller scales. The performance of SpESO is almost independent of level $j$, except for the highest level $j=14$. The figures show that SpESO has lower errors than the best $M/2$-term approximation at levels 11 to 14. LOMP is clearly not competitive compared to the other methods at any level.  It is important to remember that the good relative performance of SpESO is especially significant since the best $M/2$-term approximation requires full sampling of the signal (the nonlinear wavelet filtering is based on the full set of wavelet coefficients).

\begin{table}[tbp]
\centering\footnotesize
%\vspace{0.2cm} 
\renewcommand{\arraystretch}{1.15}
\caption{Errors in spectrum slopes of averaged estimations, $s- s^{\star}$, where $s$ is the slope of the original signal, i.e., a $k^{-s}$ power law. Slopes are computed by least-squares fitting in the range of the first slope, from $k=128$ to $k=1024$. The 1-D signal lengths are $N=2^{15}$ and the number of simulations is 64.  \label{tab:slopeerrorone}}
\begin{tabular}{cc|ccccc} 
\hline 
\rule[-0.19cm]{0cm}{0.6cm} Signal & $N/M$ & SpESO & Shannon & $M/2$-best & $M$-best & LOMP \\
\hline
         & 4 &  0.15 &  0.02 & -0.03 & -0.00 & -0.17 \\
W$(3,5/3)$ & 8 &  0.06 &  0.12 & -0.15 & -0.03 & -0.41 \\
        & 16 &  0.11 &  0.42 & -0.82 & -0.15 &  0.66 \\
\hline
         & 4 &  0.13 &  0.02 & -0.01 & -0.00 & -0.00 \\
F$(3,5/3)$ & 8 & -0.05 &  0.10 & -0.06 & -0.01 &  0.01 \\
        & 16 &  0.41 &  0.34 & -0.28 & -0.06 &  0.67 \\
\hline
         & 4 & -0.05 &  0.00 & -0.01 & -0.00 & -0.06 \\
W$(5/3,3)$ & 8 & -0.27 &  0.02 & -0.10 & -0.01 & -0.32 \\
        & 16 & -0.43 &  0.22 & -0.53 & -0.10 &  0.39 \\
\hline
         & 4 & -0.02 &  0.00 & -0.00 & -0.00 &  0.03 \\
F$(5/3,3)$ & 8 & -0.28 &  0.02 & -0.06 & -0.00 & -0.01 \\
        & 16 & -0.32 &  0.24 & -0.27 & -0.06 &  0.39 \\
\hline
\end{tabular}
\end{table}

% slope error, 1-D, signal=[w1,w2,f1,f2]
% spectra_error_1-D_plotload3.m
\begin{table}[tbp]
\centering\footnotesize
\renewcommand{\arraystretch}{1.15}
\caption{ Errors in spectrum slopes of averaged estimations, $s- s^{\star}$, where $s$ is the slope of the original signal, i.e., a $k^{-s}$ power law. Slopes are computed by least-squares fits in the range of the second slope, from $k=1024$ to $k=8192$, except for the Shannon slope which is fitted in its non-zero range only.  The 1-D signal lengths are $N=2^{15}$ and the number of simulations is 64.  \label{tab:slopeerror}}
\begin{tabular}{cc|ccccc} 
\hline 
\rule[-0.19cm]{0cm}{0.6cm} Signal & $N/M$ & SpESO & Shannon & $M/2$-best & $M$-best & LOMP \\
\hline
         & 4 & -0.06 &  0.58 & -0.78 & -0.29 &  1.46 \\
W$(3,5/3)$ & 8 & -0.50 &  1.27 & -1.92 & -0.78 &  2.23 \\
        & 16 & -1.06 & n/a & n/a & -1.92 &  2.14 \\
\hline
                 & 4 & -0.09 &  0.49 & -0.86 & -0.26 &  1.45 \\
F$(3,5/3)$ & 8 & -0.26 &  0.90 & -4.10 & -0.86 &  2.23 \\
        & 16 & -0.75 & n/a & n/a & -4.10 &  2.26 \\
\hline
                 & 4 &  0.93 &  0.54 & -1.30 & -0.33 &  2.42 \\
W$(5/3,3)$ & 8 &  0.76 &  1.65 & -4.84 & -1.30 &  3.33 \\
        & 16 &  0.25 & n/a & n/a & -4.84 &  3.28 \\
\hline
                 & 4 &  0.95 &  0.35 & -1.87 & -0.27 &  2.27 \\
F$(5/3,3)$ & 8 &  0.85 &  0.87 & -4.76 & -1.87 &  3.23 \\
        & 16 &  0.46 & n/a & n/a & -4.76 &  3.29 \\
\hline
\end{tabular}
\end{table}

Tables~\ref{tab:slopeerrorone} and \ref{tab:slopeerror} give the error of the estimates of the power law scaling of the energy spectrum over the large and small scale power law ranges (i.e. small and large wavenumber ranges).  These slopes are computed for averages of estimations.  This is a crucial quantity characterizing turbulent flows and other experimental signals. Those cases that are too bad for a reasonable fit are indicated by ``n/a''. In the range of the first power law, Table~\ref{tab:slopeerrorone}, it can be argued that SpESO, Shannon, and the $M/2$-best give on average similar results, and that the best $M$-term approximation is by far the best. SpESO performs better than the $M/2$-best term approximation in all cases, and better than the $M$-best term approximation and Shannon sampling (where it is valid) in all but two cases shown in Table~\ref{tab:slopeerror}, namely both $(5/3,3)$ cases with sampling ratio $N/M=4$. LOMP is again noticeably worse than all other methods. Even for the power law scaling at small wavenumbers, which is well-resolved by the Shannon sampling, SpESO still gives results similar to the best $M/2$ term approximation at high sampling ratios and much better than LOMP.

% E vs. k, 1-D, signal=[w1,w2;f1,f2],  best/worst cases, N/M=8
% spectra_error_1-D_plotload2.m

It is important to note that there is significant stochastic variation in the SpESO energy spectrum estimates. Figure~\ref{fig:logcases} shows the best and worst cases of SpESO as measured by the $\ell^2$ norm on a logarithmic scale when the energy spectrum slope changes at a wavenumber two times smaller than the Nyquist wavenumber. Even the worst cases are not much different from the ideal case of Shannon sampling in the low wavenumbers, although they significantly over-estimate the energy at high wavenumbers.  In addition, the worst SpESO cases seem not able to track the steepening slope at high wavenumber in the $(5/3,3)$ cases, although they do estimate approximately the correct high wavenumber slope for the $(3,5/3)$ cases (but at incorrectly high energy levels).  In contrast, the best SpESO cases estimate the spectra better than the even the best $M$-term estimates at all wavenumbers. This suggests there is potential to greatly improve the reliability and accuracy of the method if acceptable trials could be determined {\em a priori\/}.  Figure~\ref{fig:logcases2} shows the variation of the SpESO estimates in two cases where the slope of the energy spectrum changes at wavenumbers four and eight times larger than the Nyquist wavenumber.  In this case, the worst SpESO estimate is similar in accuracy to the best $M$-term approximation (both methods underestimate the energy in the second power law range), while the best SpESO result gives an excellent estimate. The averaged spectra are, however, not as responsive as the best cases.

% E vs. k, 1-D, signal=[w1;f1;w2;f2], average, N/M=[16,8,4]
% spectra_error_1-D_plotload4.m

The averages of estimated energy spectra corresponding to Tables~\ref{tab:slopeerrorone} and \ref{tab:slopeerror} are shown in Figure~\ref{fig:logspecaver3}. Comparing SpESO to Shannon, SpESO somewhat more accurately estimates the spectra beyond the Nyquist wavenumber, although performance is signal-dependent. For the $(5/3,3)$ cases, it is unclear to us why the higher ratios of $N/M$ are more accurate than the lower ratios. On average, SpESO is not worse than the best $M/2$-term approximation. 

A more severe test is to apply SpESO to signals where the change in slope is at a wavenumber higher than the equivalent Nyquist wavenumber for the sampling ratio used.  Results for this test are shown in Figure~\ref{fig:logspecaver32} when the slope changes at $k=N/32=1024$ while the equivalent Nyquist sampling wavenumber is only $k=512$.  These results show SpESO is still able to estimate the spectra for low wavenumbers, but it is not always reliable for high wavenumbers.  At these relatively large sampling ratios SpESO performs very well for the $W(3,5/3)$ cases (i.e. for intermittent cases when the slope becomes shallower at higher wavenumbers).  SpESO performs worst for the $F(5/3,3)$ cases (not shown but similar to the $W(5/3,3)$ cases) when the data is statistically non-intermittent.

% E vs. k, 1-D, signal=[w1;w2;f1;f2], average, N/M=32
% spectra_error_1-D_plotload4.m

% E vs. k, 1-D, signal=hotwire, average, N/M=[16,8,4]
% spectra_error_1-D_plotload4.m

The experiments for the hot-wire data, Figure \ref{fig:logspecaverrhh}, show the ability of SpESO to estimate the spectra beyond the capabilities of the Shannon sampling.  However, the estimates are not accurate in the range of the high wavenumber exponential decay of $E$.

\subsection{Results for 2-D Signals\label{sec:2-Dcases}} 
\begin{figure}[tbp]
\centering\footnotesize
\renewcommand{\tabcolsep}{2pt}     % width
\begin{tabular}{cccc}
\rotatebox{90}{ \qquad\qquad\qquad Fourier $(5/3)$ } &
\includegraphics[width=\figwidtht]{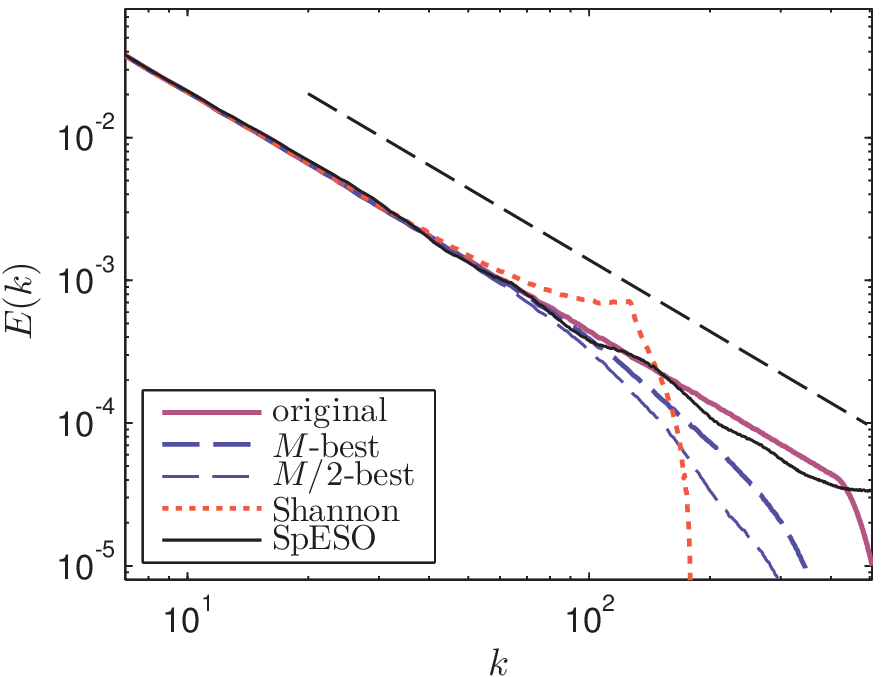}&
\includegraphics[width=\figwidtht]{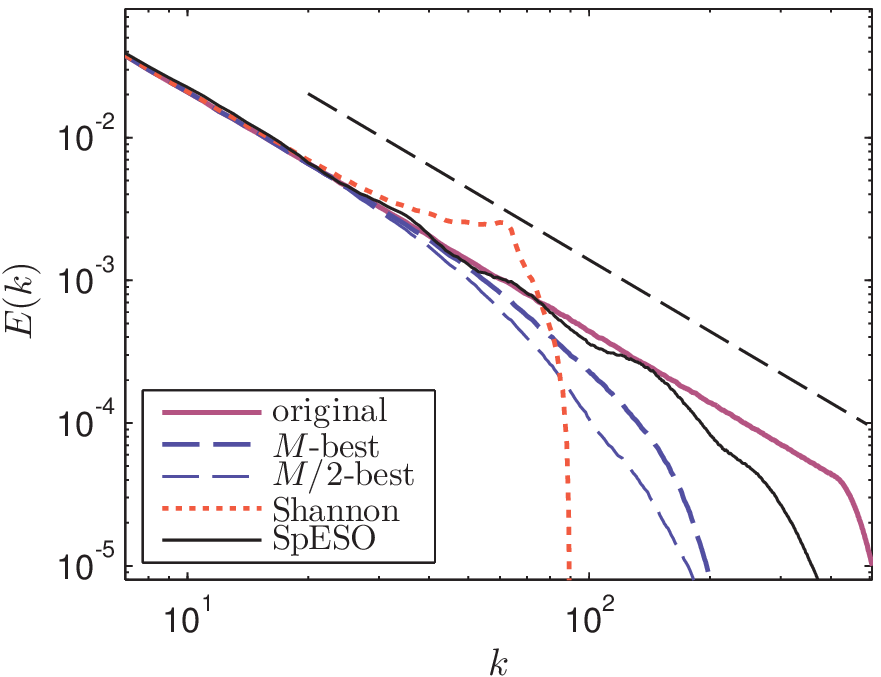}\\ 
\rotatebox{90}{ \qquad\qquad\qquad Wavelet $(5/3)$ } &
\includegraphics[width=\figwidtht]{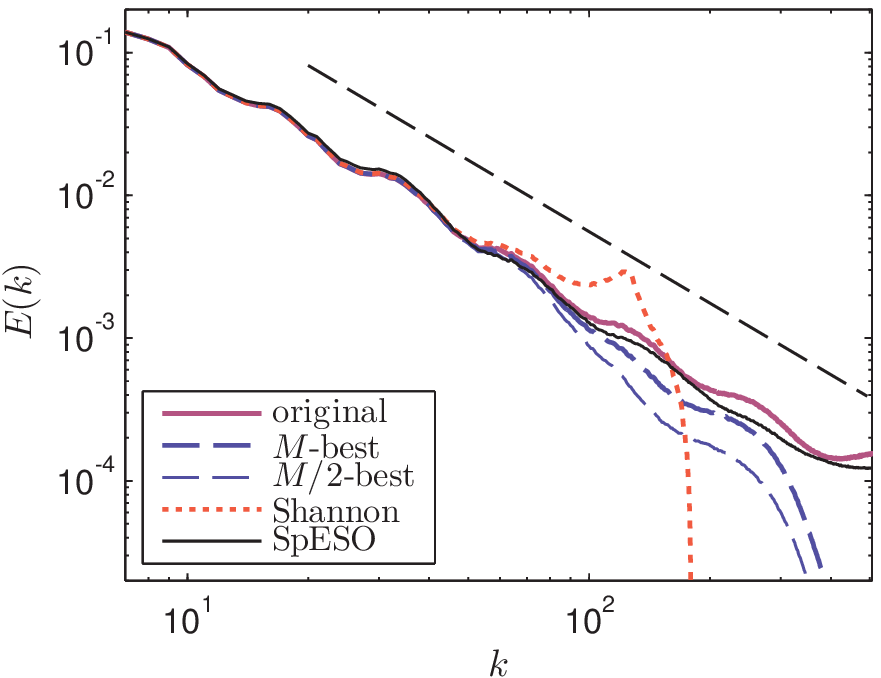}&
\includegraphics[width=\figwidtht]{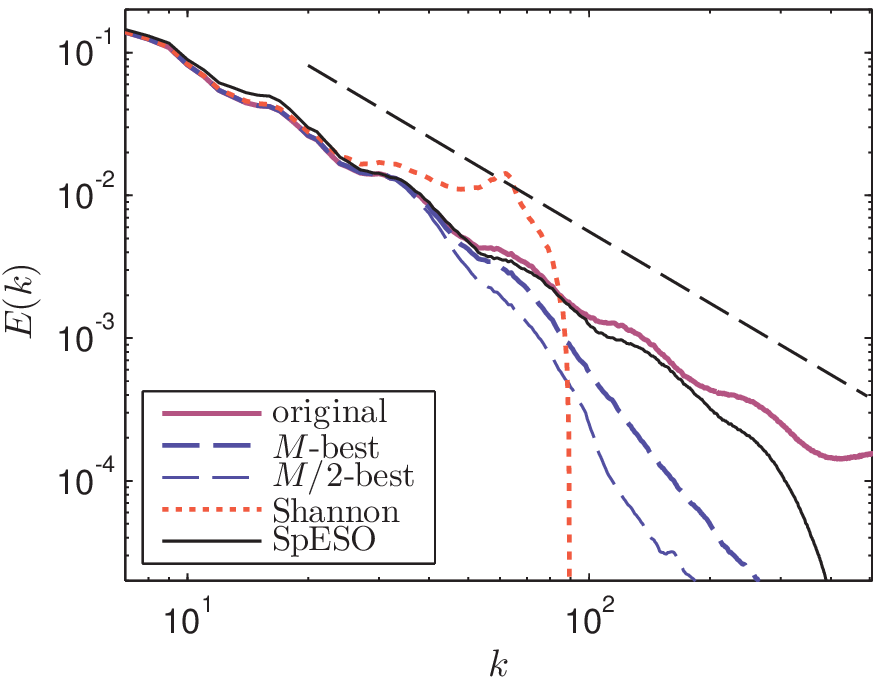}\\ 
 & $N/M=16$ & $N/M=64$ 
\end{tabular}
\caption{Logarithmic scale averages of spectrum estimates for single slope 2-D signals. The signal lengths are $N=1024^2=2^{20}$ and the number of simulations is 16.  \label{fig:log2-Dspecavers}}
\end{figure}

\begin{figure}[tbp]
\centering\footnotesize
\renewcommand{\tabcolsep}{2pt}     % width
\begin{tabular}{ccc}
\rotatebox{90}{ \qquad\qquad\qquad Fourier } &
\includegraphics[width=\figwidtht]{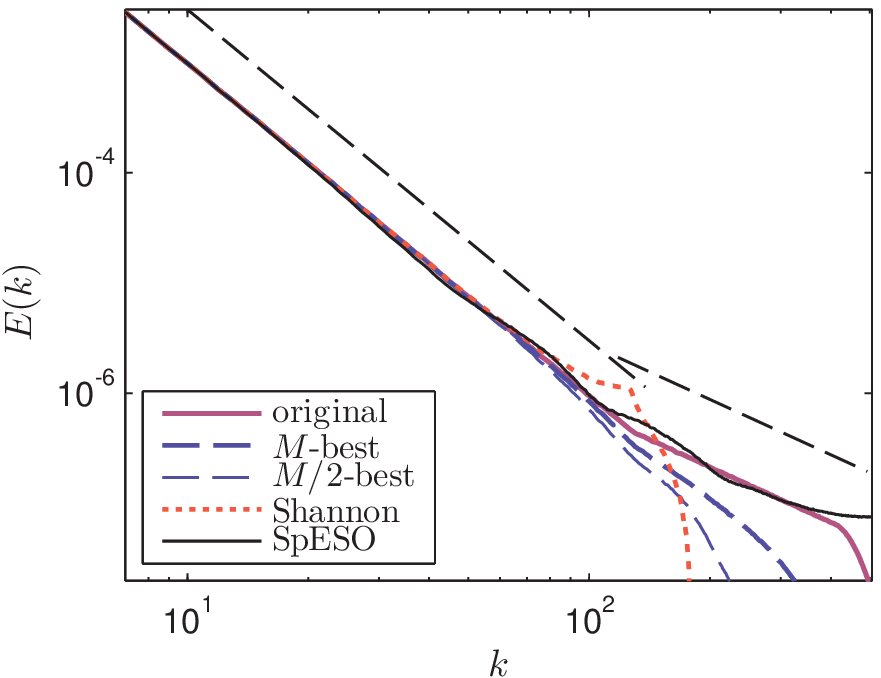}&
\includegraphics[width=\figwidtht]{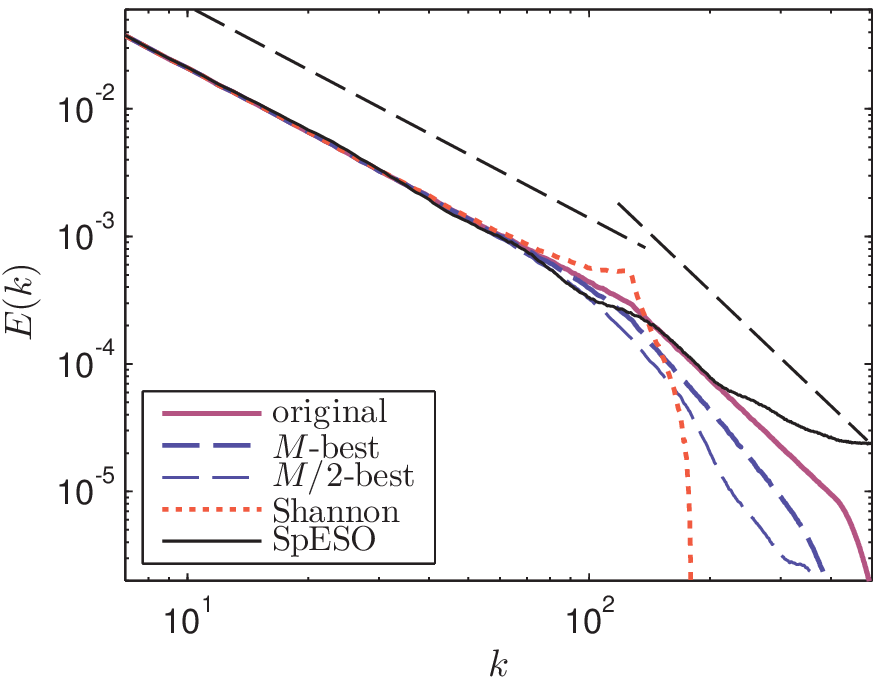}\\ 
\rotatebox{90}{ \qquad\qquad\qquad Wavelet } &
\includegraphics[width=\figwidtht]{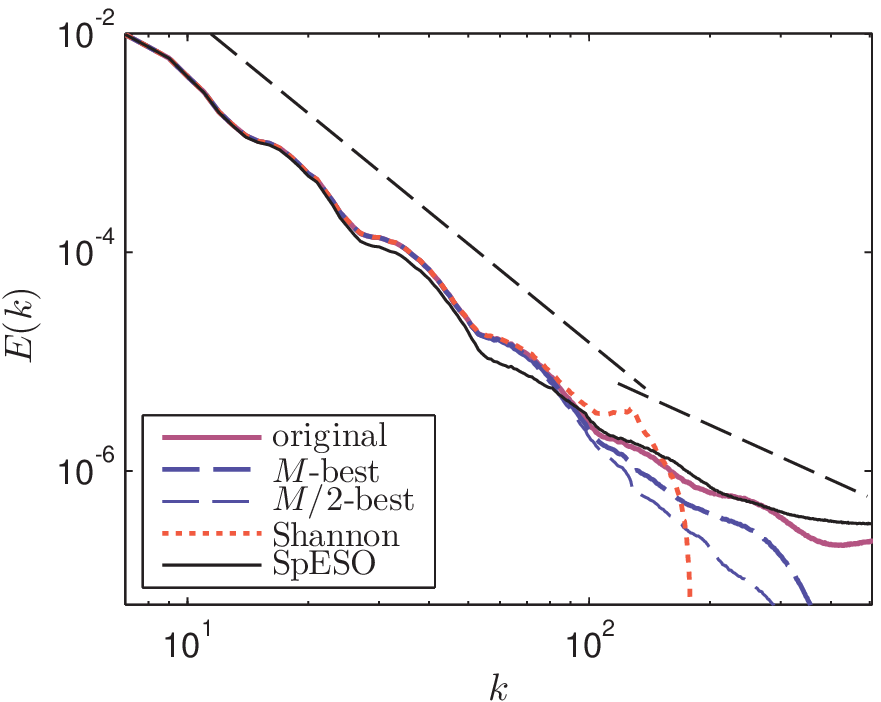}&
\includegraphics[width=\figwidtht]{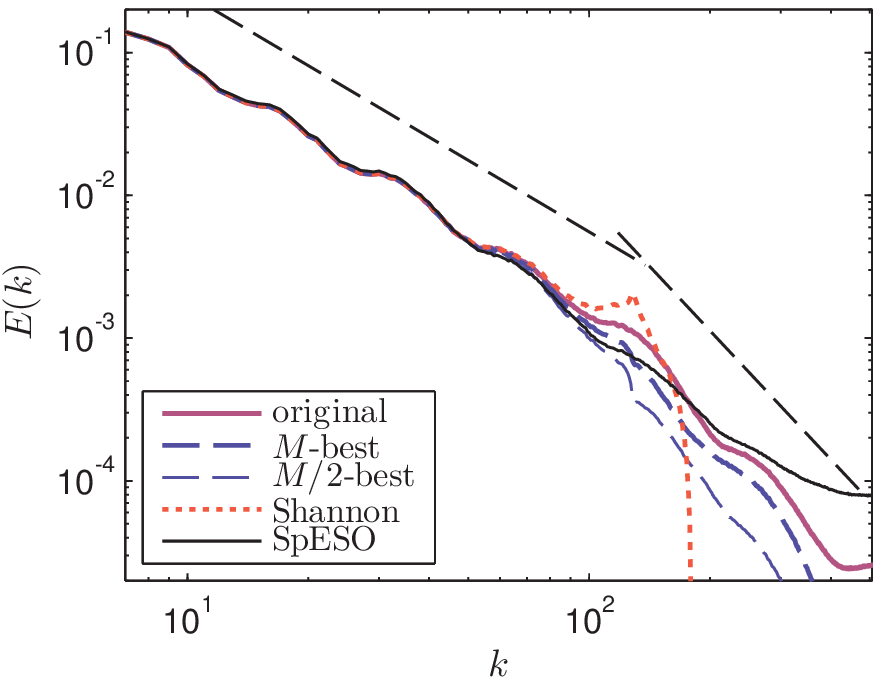}\\
 & $(3,5/3)$  &  $(5/3,3)$
\end{tabular}
\caption{ Logarithmic scale averages of spectrum estimates for 2-D signals. The signal lengths are $N=1024^2=2^{20}$, the sampling ratio is $N/M=16$, and the number of simulations is 16.  \label{fig:log2-Dav2s16}}
\end{figure}

\begin{figure}[tbp]
\centering\footnotesize
\renewcommand{\tabcolsep}{2pt}     % width
\begin{tabular}{ccc}
\rotatebox{90}{ \qquad\qquad\qquad JHU DNS } &
\includegraphics[width=\figwidtht]{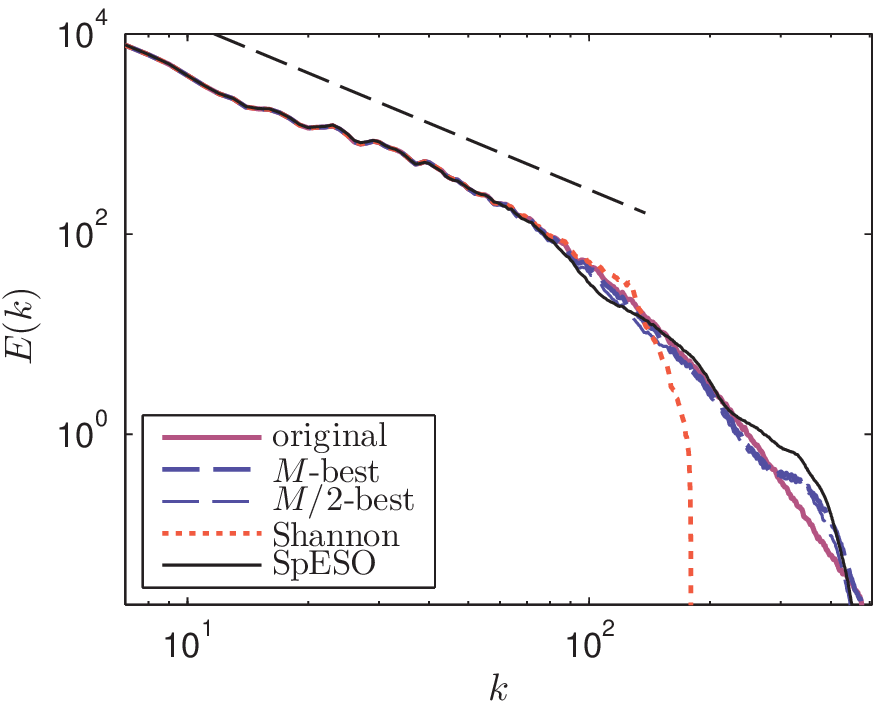}&
\includegraphics[width=\figwidtht]{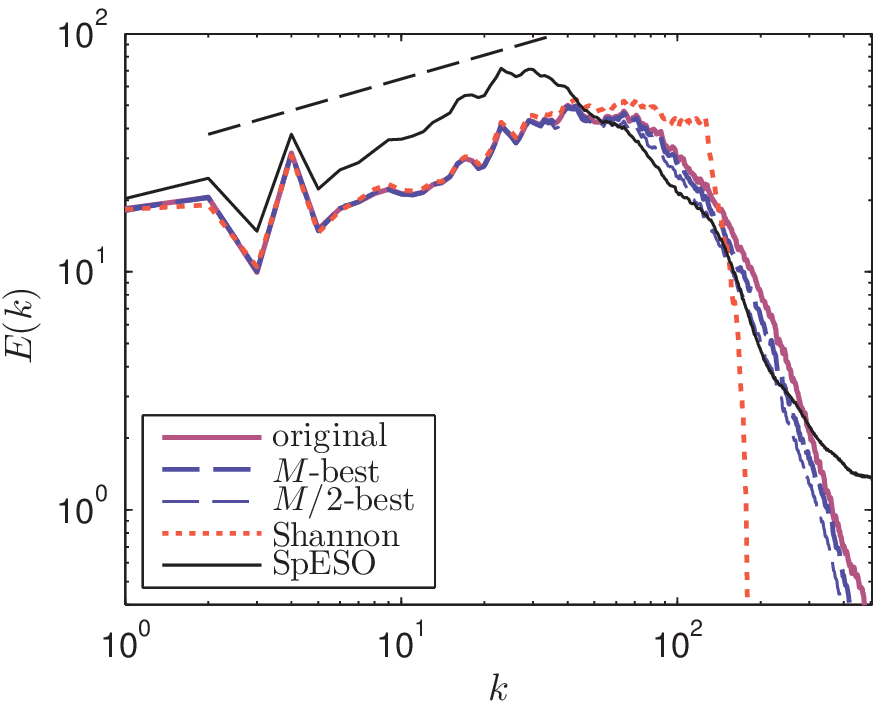}\\ 
& Velocity &  Vorticity
\end{tabular}
\caption{ Logarithmic scale averages of spectrum estimates for 2-D DNS signals. The signal lengths are $N=1024^2=2^{20}$, the sampling ratio is $N/M=16$, and the number of simulations is 16.  \label{fig:logspecaverjh}}
\end{figure}

We shall now examine 2-D signals signals of length $N=n\times n$. For indicated ratio $N/M=64$, SpESO actually has the ratio 63.3 due to the computational set-up. We note that individual 2-D estimates vary much less and are much smoother than the 1-D estimates. The results for Fourier and wavelet synthetic signals are shown in Figures~\ref{fig:log2-Dspecavers} and \ref{fig:log2-Dav2s16}, and the results for signals from the JHU DNS database are shown in Figure~\ref{fig:logspecaverjh}.

% E vs. k, 2-D, signal=[f1;w1], average, N/M=[64,16]
% spectra_error_2-D_plotload4.m

Results for single $(5/3)$ slope synthetic signals are in Figure \ref{fig:log2-Dspecavers}. Clearly, SpESO is able estimate the spectra accurately for the mid and high wavenumbers much better than Shannon sampling. The SpESO estimates are closer overall to the exact results than the best $M$-term approximations. This is surprising because the best $M$-term approximation is expected to be give the upper bound on the accuracy of SpESO since it uses all data, and then reconstructs with the best $M$ largest wavelet coefficients.  This suggests that best $M$-term approximations are not necessarily optimal for estimating nonlinear functions of the data and that carefully designed CS methods may be a better choice even if all data is available for analysis.

% E vs. k, 2-D, signal=[f2,w2;f3,w3], average, N/M=16
% spectra_error_2-D_plotload4.m

As in 1-D, we test our method for synthetic signals with a change in slope (Figure \ref{fig:log2-Dav2s16}). Apart from the wavelet $(5/3,3)$ case, the SpESO slope estimates are at least on a par with the $M/2$-best. SpESO can predict a change in a spectrum slope at the Nyquist wavenumber, which is not possible using Shannon sampling.

A significant range of the energy spectra for the JHU DNS data has an exponential decay, see Figure \ref{fig:logspecaverjh}, and in this range the best-term approximations are indeed better than SpESO---but are not far from each other in the velocity case. However, when applied to an equivalent vorticity field with a positive power law slope at small wavenumbers, SpESO captures the correct scaling, but overestimates the energy by a significant amount.

% E vs. k, 1-D, signal=[jh,jhvor], average, N/M=16
% spectra_error_2-D_plotload4.m

\section{Conclusions\label{sec:conc}}
The compressive sampling Spectrum Estimation by Sparse Optimization (SpESO) method proposed in this paper shows potential for energy spectrum estimation of signals with power law decay.  At this stage SpESO is experimental, a proof of concept, without rigorous proofs of convergence or error bounds. Nevertheless, we have derived mathematical estimates for the performance of SpESO  in section~\ref{sec:proof} and tested it numerically on a wide variety of representative synthetic, experimental, and DNS turbulence signals in one and two dimensions in section~\ref{sec:num}. 

The 2-D cases appear more promising than the 1-D cases, probably due to  the dimensionality reduction or due to different measurement matrix or both. The results show that the estimates generated by SpESO  distribute errors more evenly over the full range of wavenumbers than traditional Shannon sampling or best-term wavelet approximations. They also correctly predict the power law scaling of the energy spectrum at wavenumbers higher than those that can be captured with Shannon sampling (which is limited by the Nyquist frequency).  

Most interestingly, SpESO typically performs better than a best-term wavelet approximation using the same number of coefficients.  This is surprising because best-term wavelet approximations require the wavelet transform of the entire data set, and then select the largest $M$ terms (i.e. it uses complete information about the signal to build its approximation from a nonlinear filter of the wavelet coefficients of the data). In contrast, the SpESO method samples only $M$ data points, between 4 and 64 times fewer samples than with the best-term wavelet approximations.

Both SpESO and QOMOMP have several tuneable parameters and many possible variations. We do not suggest the method, as it is, should be immediately used in applications. However, by tuning of parameters and estimation of errors, it might be a practically useful method in those cases where high accuracy is less important than minimizing the number of samples, or where obtaining a complete set of evenly spaced measurements at the Nyquist rate is not possible.  In particular, it could be used for estimating the energy spectrum of three-dimensional or two-dimensional turbulent flows at very high Reynolds numbers where sampling at the Nyquist rate is impossible.  For example, energy spectrum estimation of atmospheric flow at a Reynolds number $\text{Re} \sim 10^{10}$ would require $\sim\! 10^{22}$ samples to fully characterize its energy spectrum if sampled in three dimensions at the Nyquist rate.  Even a 1-D measurement would require $\sim\! 10^7$ samples, which may be impractical in some cases.

It should be straightforward to extend SpESO to three dimensions, and it could be tested with measurement matrices more appropriate for field or laboratory experiments.  The same approach could be used to estimate other nonlinear functions of compressively sampled data, such as the scaling of high order exponents of turbulence structure functions $\zeta_p$, which require extremely large data sets to characterize properly for $p>10$.  SpESO could be optimized further by improving the performance of the sparsity system, for example by using wavelet packets instead of orthogonal wavelets.

This paper has shown that it is possible to design a CS-based energy spectrum estimation method that performs much better than the existing LOMP or Shannon sampling approaches, even in the case where the signal is not sparse in Fourier space.  In fact, a CS-based method can perform at least as well, and often better, than a best-term wavelet approximation that requires full sampling of the signal.

\bibliographystyle{siam} 
\bibliography{./ref/cs,./ref/turb,./ref/wavelet,./ref/books,./ref/other}{}

\end{document}